\documentclass[10pt,twoside]{article}
\usepackage{epsfig}
\usepackage{fancybox}
\usepackage{graphics}
\usepackage{amsfonts}
\usepackage{amssymb}
\usepackage{enumerate}

\newtheorem{thm}{Theorem}
\newtheorem{lemma}{Lemma}

\newtheorem{result}{Result}

\newcommand{\beq}{\begin{equation}}
\newcommand{\eeq}{\end{equation}}

\newcommand{\ie}{{\it i.e.\ }}
\newcommand{\eg}{{\it e.g.\ }}

\newcommand{\lb}{\left(}
\newcommand{\rb}{\right)}
\newcommand{\lsb}{\left[}
\newcommand{\rsb}{\right]}
\newcommand{\la}{\left\{ }
\newcommand{\ra}{\right\} }

\def\adots{\mathinner{\mskip0mu\raise0pt\vbox{\kern7pt\hbox{.}}\mskip3mu
        \raise4pt\hbox{.}\mskip3mu\raise8pt\hbox{.}\mskip0mu}}

\newcommand{\m}{{\!\, -\!\,}}
\newcommand{\p}{{\!\, +\!\,}}
\newcommand{\dsum}{\displaystyle\sum}

\newcommand{\dfrac}{\displaystyle\frac}

\newcommand{\dprod}{\displaystyle\prod}
\newcommand{\T}{H}

\newcommand{\eps}{\epsilon}

\newcommand{\rmH}{\mbox{H}}

\newcommand{\bmy}{\mbox{\boldmath $y$}}
\newcommand{\bmv}{\mbox{\boldmath $v$}}

\newcommand{\bmh}{\mbox{\boldmath $h$}}

\newcommand{\bmH}{\mbox{\boldmath $H$}}

\newcommand{\bmY}{\mbox{\boldmath $Y$}}
\newcommand{\bmV}{\mbox{\boldmath $V$}}

\newcommand{\bmX}{\mbox{\boldmath $X$}}

\newcommand{\bfH}{\mbox{\bf H}}

\newcommand{\cT}{{\mathcal T}}

\newcommand{\cN}{{\mathcal N}}
\newcommand{\cA}{{\mathcal A}}
\newcommand{\cK}{{\mathcal K}}

\newcommand{\cJ}{{\mathcal J}}

\newcommand{\cV}{{\mathcal V}}
\newcommand{\cC}{{\mathcal C}}

\newcommand{\cM}{{\mathcal M}}

\newcommand{\Lu}{\underline{M}}
\newcommand{\E}{\mbox{E}}
\newcommand{\rank}{\mbox{rank}}
\newcommand{\trace}{\mbox{tr}}
\newcommand{\re}{\mbox{Re}}
\newcommand{\im}{\mbox{Im}}
\newcommand{\thetah}{\hat{\theta}}

\newcommand{\thetat}{\tilde{\theta}}

\newcommand{\ds}{\displaystyle}

\newcommand{\hh}{\hat{{h}}}
\newcommand{\hhh}{\hat{\hat{h}}}

\newcommand{\Vec}{\mbox{vec}}

\newcommand{\range}{\mbox{range}}
\newcommand{\sv}{{\sigma_v^2}}
\newcommand{\sa}{{\sigma_a^2}}

\newcommand{\Rr}{\mathbb{R}}
\newcommand{\Cr}{\mathbb{C}}

\renewcommand{\thefootnote}{\fnsymbol{footnote}}
\begin{document}

\bibliographystyle{IEEEbib}
\thispagestyle{empty}

\begin{center}
{\Large\bf
Cram\'er--Rao Bounds
\vspace{4mm} \\
for Blind Multichannel Estimation\footnote{
The work of Elisabeth de Carvalho was supported by Laboratoires
d'Electronique Philips under contract Cifre 297/95.
The work of Dirk Slock was supported in part by Laboratoires 
d'Electronique Philips under contract LEP 95FAR008.}}
\vspace{7mm}\\
\begin{minipage}[t]{2.7in}
\begin{center}
Elisabeth de Carvalho\\[2mm]
Aalborg University\\
Dept. of Electronic Systems\\
Selma Lagerl{\O}fs Vej 312\\
9220 Aalborg, DENMARK\\
Tel: +45 9940 7528\\
Email: edc@es.aau.dk
\end{center}
\end{minipage} \ and \ 
\begin{minipage}[t]{2.7in}
\begin{center}
Dirk T.M.\  Slock\\[2mm]
EURECOM\\
Communication Systems Dept.\\
450 Route des Chappes\\
06410 BIOT Sophia Antipolis, FRANCE\\
Tel: +33 4 93 00 26 06\\
email: slock@eurecom.fr
\end{center}
\end{minipage}
\vspace{5mm}\\
Written December 1999
\vspace{5mm}\\
\end{center}

\renewcommand{\thefootnote}{\arabic{footnote}}
\setcounter{footnote}{0}

\begin{abstract}
In some estimation problems, not all the parameters can be identified,
which results in singularity of the  Fisher Information Matrix (FIM). 
The  Cram\'er--Rao Bound (CRB), which is the inverse of the FIM, is then not defined.
To regularize the estimation problem, one can impose constraints on the parameters
and derive the corresponding CRBs. The correspondence between 
local identifiability and FIM regularity is studied here. 
Furthermore  the  number of FIM singularities
is shown to be equal to the number of independent  constraints 
necessary to have a well--defined
constrained CRB and local identifiability. 
In general, many sets of constraints can render the parameters 
identifiable, giving different values for the CRB, that are not always
relevant. When  the constraints can be chosen, we 
propose a constrained CRB, the pseudo-inverse of the FIM, which 
gives, for a minimum number of constraints, the
lowest bound on the mean squared estimation error.
These results are applied to two approaches to blind FIR multichannel estimation
which allow identification of the channel up to a scale or phase factor.
These two approaches correspond to deterministic and Gaussian models for the
unknown channel inputs.  
The singularities of the FIMs and local identifiability are studied and the
corresponding constrained CRBs are derived and interpreted. 

\end{abstract}

\vspace{3mm}

\begin{center}
\begin{tabular}{ll}
Keywords: & Fisher Information Matrix, Cram\'er--Rao Bound,  \\[0mm]
 &  constrained parameter estimation, blind channel  identification. 
\end{tabular}
\end{center}

%%%%%%%%%%%%%%%%%%%%%%%%%%%%%%%%%%%%%%%%%%%%%%%%%%%%%%%%%%%%%%%%%%%%%%%%%%%
%%%%%%%%%%%%%%%%%%%%%%%%%%%%%%%%%%%%%%%%%%%%%%%%%%%%%%%%%%%%%%%%%%%%%%%%%%%
\newpage

\pagenumbering{arabic}

%%%%%%%%%%%%%%%%%%%%%%%%%%%%%%%%%%%%%%%%%%%%%%
\section{Introduction}
%%%%%%%%%%%%%%%%%%%%%%%%%%%%%%%%%%%%%%%%%%%%%%

The Cram\'er-Rao Bound (CRB) is a powerful tool in estimation theory
as it gives a performance lower bound for parameter estimation problems.
It is computed as the inverse of the Fisher Information Matrix (FIM). 
When the parameters cannot be completely identified, the FIM is 
singular, and the classical CRB results cannot be applied directly. 

The main underlying motivation of this work is the study of the performance of 
certain blind channel estimation problems where the parameters can indeed be 
identified only up to scale or phase factor~\cite{carvalho:Ident99}.
Blind estimation is done under certain parameter constraints to 
regularize the problem. The  performance of blind methods is not correctly 
evaluated in general or remains somewhat vague. A constraint often 
used~\cite{Moulines:sp95}
is to consider one coefficient of the channel as known (which is sufficient 
to render the estimation problem regular): the 
resulting performance and its bound depend on the choice of this coefficient and appear
arbitrary. One of the contributions of this work will be to give 
a less arbitrary bound and the corresponding set of constraints.
Another motivation comes from other work~\cite{carvalho:Ident99,carvalho:SBCRB99}
by the  authors  in which  blind and semi--blind methods are compared  through the
CRBs. To get a meaningful comparison, semi--blind and blind CRBs have to be 
computed under the same constraints. For that purpose, 
this study, which is  valid for the regular
or the non regular estimation problem, was then necessary.

The first part of this paper focuses first on the FIMs and especially the 
correspondence, for a Gaussian data distribution, between FIM regularity 
and parameter identifiability, defined in terms of a probability density 
function. 
%Under a certain condition, FIM regularity and local identifiability  
%are equivalent in general. For the blind channel estimation application
%considered here, this equivalence is true without any extra conditions. 
For the blind channel estimation application considered here,
FIM regularity and local identifiability  are equivalent.
In a second step, we study the CRBs for estimation under parameter constraints. 
A similar study was done in~\cite{Hero:it90} for
the case where the unconstrained problem is identifiable, \ie the FIM is regular. 
We adapt here the results to the case where the unconstrained problem 
leads to nonidentifiability, \ie the FIM is singular. 
We furthermore outline the correspondence between the number and characteristics
of FIM singularities and the number and characteristics of independent constraints
needed in order to regularize the estimation problem and to be able 
to define the constrained CRB. 
In a last step, assuming that we can choose the set of constraints, 
we propose a particular CRB for the case of an unidentifiable  unconstrained estimation
problem: this CRB is the Moore-Penrose pseudo-inverse of the 
FIM. It corresponds to a minimum number of particular constraints 
and gives the lowest bound on the mean square estimation error, 
\ie $\trace(CRB)$.

The second part of this paper focuses on two classes of blind FIR multichannel 
estimation problems corresponding to two different models for the input symbols. 
The deterministic model, which exploits no statistical information
on the input symbols,  takes the input symbols to be deterministic quantities
whereas in the Gaussian model we consider them to be Gaussian random variables
to exploit their second--order statistics. 
The deterministic model leads  to the class of methods directly based
on the structure of the received signal; the Gaussian approach includes methods 
based on the second--order moments of the data, like certain prediction 
approaches~\cite{AbedMeraim:sp97} or the covariance matching method~\cite{giannakis:sp95}.
We refer to~\cite{carvalho:Ident99}  for a more complete description 
of the two models. The deterministic methods can identify the channel up to a
scale factor only and the Gaussian methods up to a phase factor, resulting 
in singularities of the FIM.

The FIM singularities are studied in both cases and hence also the conditions for
local identifiability. The deterministic model cannot identify the zeros of a
multichannel whereas the Gaussian model can, under certain conditions given
here. 
Throughout the paper, 
we distinguish between the real and complex parameter cases since they lead to 
different FIMs,  with different singularities, 
and require different  regularization constraints.
The blind deterministic CRB is computed under the commonly used 
norm constraint which imposes the norm of the channel to be constant.
This constraint is  sufficient to
regularize the problem when the channel is real, but not when it is complex in which case an additional 
constraint is required to adjust the phase of the channel. This constraint is
chosen so that the resulting  constrained CRB is the Moore-Penrose
pseudo--inverse of the FIM and  corresponds to a minimal
constrained  CRB. When the channel is real the Gaussian FIM is regular, 
when it is complex however, the FIM is singular: a constraint on the 
phase is necessary as in the deterministic case and
the constrained CRB is again the pseudo--inverse of the FIM.
In some examples, we illustrate furthermore the variation of the constrained
CRB according to the constraint chosen and especially compare the
pseudo--inverse of the FIM to the constrained CRB corresponding to the 
assumed knowledge of one coefficient of the channel.

%%%%%%%%%%%%%%%%%%%%%%%%%%%%%%%%%%%%%%%%%%%%%%%%%%%%%%%%%%%%%%%

Some notations and acronyms that will be used in the paper are:
\begin{tabbing}
\hspace{6mm} \= $(.)^*$, $(.)^T$,  $(.)^H$  \hspace{6mm} \= conjugate,
transpose, conjugate transpose \\
\> $(.)^+$ \> Moore--Penrose pseudo--inverse \\
\> $\trace(A)$, $\det(A)$ \> trace and determinant of matrix $A$\\
\> $\Vec(A)$ \> $[A^T_{i,1} \;\; A^T_{i,2}  \cdots A^T_{i,n}]^T $\\
\> $\otimes$ \> Kronecker product\\
\> $\thetah$, $\theta^o$ \> estimate of parameter $\theta$, true value of parameter $\theta$ \\
\> $\E_X$ \> mathematical expectation w.r.t.\ the random quantity $X$ \\
\> $\re(.)$, $\im(.)$  \> real and imaginary part\\
%\> $A_{n,m}$ \> matrix $A$ of dimension $n \times{}m$ (?)\\
\> $I$ \> identity matrix with adequate dimension \\
%\> SOS \> second Order Statistics\\
\> w.r.t.\ \> with respect to \\
\end{tabbing}

%%%%%%%%%%%%%%%%%%%%%%%%%%%%%%%%%%%%%%%%%%%%%%
\section{CRBs for Real and Complex Parameters}
%%%%%%%%%%%%%%%%%%%%%%%%%%%%%%%%%%%%%%%%%%%%%%

We assume here the FIMs to be regular. 

%------------------------------------------
\subsection{CRBs for Real Parameters}
%------------------------------------------

Let $\theta$ be a deterministic  real parameter vector and $f(\bmY|\theta)$ the
probability density function of the vector of observations $\bmY$. 
The FIM associated with $\theta$ is:
\beq
\cJ_{\theta\theta}=
\E_{Y|\theta} \lb \dfrac{\partial \ln f(\bmY|\theta)}{\partial \theta}\rb \lb
\dfrac{\partial \ln f(\bmY|\theta)}{\partial \theta}\rb^T \; . 
\label{C2eq1}
\eeq
Let $\thetah$ be an unbiased estimate of $\theta$ and 
$\thetat=\thetah-\theta$ the estimation error. Hence $\E \thetat=0$ and 
$\cC_{\thetat\thetat}=\E\thetat\thetat^T$ is the 
error covariance matrix. 
When $\cJ_{\theta\theta}$ is nonsingular and 
under certain regularity conditions~\cite{Kay:book}, $\cJ_{\theta\theta}^{-1}$ is the 
Cram\'er--Rao Bound and:
\beq
\cC_{\thetat\thetat} \geq CRB = \cJ^{-1}_{\theta\theta} \; .
\label{C2eq2}
\eeq
Equality is achieved if and only if:
\beq
\thetah - \theta = 
\cJ^{-1}_{\theta\theta}
\dfrac{\partial \ln f(\bmY|\theta)}{\partial \theta} \; .
\label{C2eq3}
\eeq

%------------------------------------------
\subsection{CRB for Complex Parameters, Complex CRB.}
%------------------------------------------

When $\theta$ is a complex deterministic parameter, 
the previous results can be applied to 
$\theta_R=\lsb \re^T(\theta) \;\; \im^T(\theta)\rsb^T$ 
and $\bmY_R=\lsb \re^T(\bmY) \;\; \im^T(\bmY) \rsb^T$,
the associated real parameters
and real observations.

It is however possible to define the FIM for $\theta_R$ 
w.r.t. complex FIM--like matrices.
Let $J_{\varphi \psi}$ be defined as:
\beq
J_{\varphi \psi}=\E_{Y|\theta} \lb \dfrac{\partial \ln f(\bmY|\theta)}{\partial \varphi^*}\rb \lb \dfrac{\partial \ln f(\bmY|\theta)}{\partial \psi^*}\rb^H 
\label{C2eq4}
\eeq
where $f(\bmY|\theta)= f(\bmY_R|\theta)=f(\bmY_R|\theta_R)$. 
Derivation w.r.t.\ the  complex vector $\theta=\alpha+j\beta$ is defined as:
$
\dfrac{\partial{}}{\partial{\theta}}=
\dfrac{1}{2}
\lb
\dfrac{\partial{}}{\partial{\alpha}}
- j
\dfrac{\partial{}}{\partial{\beta}}
\rb$ 
(see~\cite{Kay:book} for some properties of complex derivation).
Remark that we denote real and complex FIMs by $\cJ$ and $J$ respectively. 

The parametrization in $(\re(\theta), \; \im(\theta))$ is equivalent to a
parametrization in terms of $(\theta, \; \theta^*)$ via:
\beq
\theta_R = 
\lsb 
\begin{array}{c}
\re(\theta) \\
\im(\theta)
\end{array}
\rsb 
 = 
\cM 
\lsb 
\begin{array}{c}
\theta \\
\theta^*
\end{array}
\rsb,
\hspace{4mm} 
\cM=
\dfrac{1}{2}
 \lsb 
\begin{array}{cc}
I & I \\
-j I & j I
\end{array}
\rsb 
\label{C2eq5}
\eeq
where $\cM$ is non--singular. 
Knowing that $J_{\theta \theta}=J^*_{\theta^*\theta^*}$
and  $J_{\theta\theta^*}=J^*_{\theta^*\theta}$  (true here as
$f_{\bmY|\theta}(\theta,\theta^*)=f_{\bmY|\theta}(\theta^*,\theta)$),
equation (\ref{C2eq5}) implies: 
\beq
\cJ_{\theta_R\theta_R} = 
\cM 
\lsb 
\begin{array}{cc}
J_{\theta\theta} & J_{\theta\theta^*} \\
J^*_{\theta\theta^*} & J^*_{\theta\theta}
\end{array}
\rsb 
\cM^H.
\label{C2eq6}
\eeq 
$\cJ_{\theta_R\theta_R}$ can be determined from 
$J_{\theta \theta}$ and  $J_{\theta\theta^*}$  as follows:
\beq
\cJ_{\theta_R\theta_R}=
2
\lsb
\begin{array}{lr}
{\mbox Re}(J_{\theta \theta})  & \m{\mbox Im}(J_{\theta \theta}) \\
{\mbox Im}(J_{\theta \theta}) & {\mbox Re}(J_{\theta \theta})
\end{array}
\rsb
+
2
\lsb
\begin{array}{lr}
{\mbox Re}(J_{\theta \theta^*}) & {\mbox Im}(J_{\theta \theta^*}) \\
{\mbox Im}(J_{\theta \theta^*}) & \m{\mbox Re}(J_{\theta \theta^*})
\end{array}
\rsb \; .
\label{C2eq7}
\eeq
To quantify the estimation  quality, the  quantity of interest will be
$\trace(CRB_R)$, \ie the lower bound on the mean squared estimation error, which can be expressed 
directly in terms of  the quantities $J_{\theta\theta}$ 
and $J_{\theta\theta^*}$. Since $\cM \cM^H = \dfrac{1}{2}I$, (\ref{C2eq6}) implies: 
\beq 
\cJ^{-1}_{\theta_R\theta_R} = 
4\;
\cM 
\lsb 
\begin{array}{cc}
J_{\theta\theta} & J_{\theta\theta^*} \\
J^*_{\theta\theta^*} & J^*_{\theta\theta}
\end{array}
\rsb^{-1}
\cM^H\; .
\label{C2eq8}
\eeq
Then: 
\beq
\trace(CRB_R) = 
\trace(\cJ^{-1}_{\theta_R\theta_R}) = 4\; \trace\lb J_{\theta\theta} - 
J_{\theta\theta^*} J^{\m *}_{\theta\theta} J^*_{\theta\theta^*} \rb^{-1}\; .
\label{C2eq9}
\eeq

\begin{thm}
When $J_{\theta \theta^*}=0$,
$\cJ_{\theta_R\theta_R}$ is completely determined 
by $J_{\theta \theta}$. In that case, $J_{\theta \theta}$ can be considered as
the {\it complex FIM} and the corresponding {\it complex CRB}
%, denoted simply as $CRB$, 
is such that:
\beq
C_{\thetat\thetat}=\E\thetat \thetat^H \geq CRB = J^{-1}_{\theta \theta}\; .
\label{C2eq10}
\eeq
If $J_{\theta \theta^*} \neq 0$, $J^{-1}_{\theta \theta}$ is also 
a lower bound on $C_{\thetat\thetat}$, but not as tight as the (real) CRB, $CRB_R$.
\end{thm}

In that  case ($J_{\theta\theta^*}=0$), a single complex singularity 
of the complex FIM $J_{\theta\theta}$ corresponds to two
real singularities since if $\theta_s$ is a singular vector, 
then so is $j \theta_s$.

%%%%%%%%%%%%%%%%%%%%%%%%%%%%%%%%%%%%%%%%%%%%%%
\section{CRBs for a Gaussian Data Distribution}
%%%%%%%%%%%%%%%%%%%%%%%%%%%%%%%%%%%%%%%%%%%%%%

%------------------------------------------
\subsection{Real Parameters}
%------------------------------------------

The CRB for a Gaussian data distribution:
\beq
\begin{array}{ll}
\bmY \sim \cN(m_Y(\theta), C_{YY}(\theta)), \hspace{5mm} 
&  m_Y(\theta)  =  \E_{Y|\theta}\lb\bmY\rb\\
&  C_{YY}(\theta) =  \E_{Y|\theta} \lb \bmY - m_Y(\theta)\rb \lb
\bmY-m_Y(\theta)\rb^H
\end{array}
\label{C2eq11}
\eeq
is~\cite{Kay:book}:
\beq
\cJ_{\theta\theta}(i,j) =
\lb \dfrac{\partial m^T_{Y}}{\partial \theta_i} \rb C^{-1}_{YY}
\lb \dfrac{\partial m^T_{Y}}{\partial \theta_j} \rb^T
+
\dfrac{1}{2} 
\trace 
\la 
C^{-1}_{YY} \lb \dfrac{\partial C_{YY}}{\partial \theta_i}\rb C^{-1}_{YY} \lb \dfrac{\partial C_{YY}}{\partial \theta_j}\rb
\ra,
\label{C2eq12}
\eeq
where, to simplify, the mean and the covariance matrix are denoted $m_Y$ and $C_{YY}$.

The FIM can also be expressed in a closed form.
Let's define $\phi$ to be a vector comprising the elements of the mean and covariance
of the data as:
\beq
\phi = 
\lsb
\begin{array}{c}
m_{Y} \\
\Vec\{C_{YY}\}
\end{array}
\rsb\; .
\label{C2eq13}
\eeq
Using the properties: $\trace\{AB\}=\Vec^T\{A^T\}\Vec\{B\}$ and $\Vec\{ABC\} =
(C^T \otimes A) \Vec\{B\}$, we find:
\beq
FIM = 
\lb \dfrac{\partial m^T_{Y}}{\partial \theta}  \rb 
C^{-1}_{YY}  
\lb \dfrac{\partial m^T_{Y}}{\partial \theta}  \rb^T
+
\lb \dfrac{\partial \Vec^T \!\{C_{YY}\}}{\partial \theta}  \rb 
\lb C^{-T}_{YY} \otimes C^{-1}_{YY} \rb 
\lb \dfrac{\partial \Vec^T \!\{C_{YY}\}}{\partial \theta}  \rb^T
\label{C2eq14}
\eeq
or
\beq
J_{\theta\theta} = 
\lb \dfrac{\partial \phi^T}{\partial \theta}\rb
\lsb 
\begin{array}{cc}
 C^{-1}_{YY}  & 0 \\ 0 & C^{-T}_{YY} \otimes C^{-1}_{YY}
\end{array}
\rsb
\lb \dfrac{\partial \phi^T}{\partial \theta}\rb^T \; .
\label{C2eq15}
\eeq
From this expression, the following theorem, also given in~\cite{Caines:book}, is deduced:

\begin{thm}
The FIM for a Gaussian data distribution is regular if and only if 
$\lb \dfrac{\partial \phi^T}{\partial \theta}\rb$ has full row rank.
\end{thm}

%(((This result shows that the study on FIM regularities that will follow
%is not only valid for Gaussian data
%distribution, is is also valid for all the methods which are based on the first
%and second--order moments of the data. The FIM depends indeed only on the
%second--order moments of the data. )))

%------------------------------------------
\subsection{Complex Parameters}
%------------------------------------------

In a properly formulated blind channel estimation problem, 
$\bmY$ and $\theta$ are simultaneously
real or complex. If $\bmY$ is complex, we shall assume it is circular, \ie
$\E\bmY \bmY^T=0$. In that case, 
it is possible to define a complex Gaussian 
conditional probability  density function for $\bmY$: 
\beq
f(\bmY|\theta)=\dfrac{1}{\pi^{Mm} \det C_{YY}(\theta)} \exp \lsb
 -\lb \bmY - m_Y(\theta)\rb^H C_{YY}^{-1}(\theta) \lb \bmY - m_Y(\theta) \rb
\rsb
\label{C2eq16}
\eeq
where
$m_Y(\theta)=\E_{Y|\theta}\lb\bmY\rb$ and 
$C_{YY}(\theta)=\E_{Y|\theta} \lb \bmY - m_Y(\theta)\rb \lb
\bmY-m_Y(\theta)\rb^H$. 
Based on the complex probability  density function $f(\bmY|\theta)$, 
the computation of 
the  complex FIMs $J_{\theta\theta}$ and $J_{\theta\theta^*}$ gives
(straightforward extension of~\cite{Kay:book}):
\beq
\la
\begin{array}{l}
J_{\theta\theta}(i,j) =
\lb \dfrac{\partial m^H_{Y}}{\partial \theta_i^*} \rb C^{-1}_{YY}
\lb \dfrac{\partial m^H_{Y}}{\partial \theta_j^*} \rb^H
+
\trace 
\la 
C^{-1}_{YY} \lb \dfrac{\partial C_{YY}}{\partial \theta_i^*}\rb  C^{-1}_{YY}
\lb  \dfrac{\partial C_{YY}}{\partial \theta_j^*}\rb^{H}
\ra \\ [4mm]
J_{\theta\theta^*}(i,j)  =
\lb \dfrac{\partial m^H_{Y}}{\partial \theta_i^*} \rb C^{-1}_{YY}
\lb \dfrac{\partial m^H_{Y}}{\partial \theta_j^*} \rb^T
+
\trace 
\la 
C^{-1}_{YY} \lb \dfrac{\partial C_{YY}}{\partial \theta_i^*}\rb C^{-1}_{YY} \lb \dfrac{\partial C_{YY}}{\partial \theta_j^*}\rb
\ra\; .
\end{array}
\right.
 \label{C2eq17}
\eeq
Proceeding as in the real case, the FIM for $\theta_R$ becomes:
\beq
\cJ_{\theta_R\theta_R}=
2 \cM
\lsb
\begin{array}{c}
\dfrac{\partial \phi^T}{\partial \theta^*} \\ \dfrac{\partial \phi^T}{\partial
  \theta}  
\end{array}
\rsb
\lb I_2 \otimes
\lsb 
\begin{array}{cc}
C^{-1}_{YY} & 0 \\ 0 &  C^{-T}_{YY} \otimes C^{-1}_{YY}
\end{array}
\rsb
\rb
\lsb
\begin{array}{c}
\dfrac{\partial \phi^T}{\partial \theta^*} \\ \dfrac{\partial \phi^T}{\partial
  \theta}  
\end{array}
\rsb^H
\cM^H\; .
\label{C2eq18}
\eeq

\begin{thm}
The FIM for a complex Gaussian data distribution is regular if and only if 
$\lsb \begin{array}{c}
 \dfrac{\partial \phi^T}{\partial \theta^*} \\
\dfrac{\partial \phi^T}{\partial \theta}
\end{array}
\rsb
$ has full row rank.
\end{thm}

%%%%%%%%%%%%%%%%%%%%%%%%%%%%%%%%%%%%%%%%%%%%%%
\section{Correspondence between Identifiability and  FIM Regularity for a Gaussian
Data Distribution}
%%%%%%%%%%%%%%%%%%%%%%%%%%%%%%%%%%%%%%%%%%%%%%
\label{section41}

\subsection{Regular Estimation}

We consider identifiability as defined in~\cite{hochwald:cssp97,carvalho:Ident99}.
In the regular case, $\theta$ is said identifiable if: 
\beq
 \forall \mbox{ } \bmY, \hspace{3mm} f(\bmY|\theta)=f(\bmY|\theta')
\hspace{3mm} \Rightarrow \hspace{3mm}
\theta=\theta'\; .
\label{C2eq19}
\eeq
When the observations have a normal distribution, $\bmY\sim\cN(m_{Y}(\theta),C_{YY}(\theta))$,
identifiability is based on the mean and covariance:
$\theta$ is said identifiable if: 
\beq
m_{Y}(\theta)=m_{Y}(\theta') \mbox{ and } C_{YY}(\theta)=C_{YY}(\theta')
\hspace{3mm} \Rightarrow \hspace{3mm}
\theta=\theta'.
\label{C2eq20}
\eeq
(the mapping: $\theta \rightarrow \{m_Y, C_{YY}\}$ is invertible). We have local
identifiability at $\theta$ if identifiability holds for $\theta'$ being
restricted to some open neighborhood of $\theta$.
In the Gaussian distribution case, there is a  correspondence between FIM regularity and local
identifiability. 

\begin{thm}
If $\theta$ is not locally identifiable at $\theta$,  
then the FIM is singular at $\theta$~\cite{Caines:book}. 
\end{thm}

If a parameter can only be identified up to a continuous ambiguity, for example a scale
factor, it cannot be locally identifiable and the corresponding FIM is 
singular.
However, when the parameter is identifiable up to a discrete
ambiguity, like a sign factor  for example,  
local identifiability holds, and the FIM can be non--singular.
Under weak conditions, local identifiability implies  FIM regularity~\cite{Caines:book}:
\begin{thm}
Assume that the  FIM is of constant rank in the neighborhood of $\theta$.  If
$\theta$ is locally identifiable, then the FIM is regular at $\theta$. 
\end{thm}
And so we have the following theorem:
\begin{thm}
Assume that the  FIM is locally of constant rank at point $\theta$, then 
$\theta$ is locally identifiable if and only if the FIM is regular at $\theta$. 
\end{thm}
For the blind channel estimation problem, we shall  
see (sections~\ref{secdet} and \ref{secgaus})
that this equivalence holds without the local rank assumption for the FIMs.

\subsection{Blind Estimation}

Blind estimation (to be introduced further below) does not allow
to  identify the parameters. They are at best
identifiable up to a certain blind ambiguity, a scale or a phase factor in our 
examples, which will be continuous in general, resulting in a singular FIM.  
Blind identifiability~\cite{carvalho:Ident99} is defined as in the regular case
except that the condition is for the parameters to be identifiable up to the
blind ambiguity. 

In the deterministic and Gaussian  input cases, local blind identifiability will be  guaranteed if and only
if the FIM has as many singularities as the number of continuous blind
ambiguities, see Table~1.
\begin{table}[h]
\begin{center}
\begin{tabular}{|c||c|c|} \hline
Number of  Real & Deterministic  & Gaussian \\
Continuous Ambiguities & Input & Input \\ \hline \hline
  Real Data &  1  & 0 \\  \hline
Complex Data  & 2 & 1 \\   \hline
\end{tabular}
\end{center}
\vspace{-3mm}
\caption{Number of Continuous Ambiguities for the Deterministic and Gaussian Model}
\end{table}
%(\ie the scale factor in the deterministic model and the phase
%factor in the Gaussian model).
Furthermore, there will be as many independent constraints needed as the number
of singularities to regularize the estimation problem.

%%%%%%%%%%%%%%%%%%%%%%%%%%%%%%%%%%%%%%%%%%%%%%
\section{CRBs for Estimation with Constraints}
%%%%%%%%%%%%%%%%%%%%%%%%%%%%%%%%%%%%%%%%%%%%%%
\label{sectionconst}

In this section, we consider real parameters (hence $\theta$ stands for
$\theta_R$ if $\theta$ is complex).
When the estimation is (locally) unidentifiable, the FIM is singular and the 
classical CRB result (\ref{C2eq2}) cannot be applied; \eg
the channel cannot be estimated by blind estimation and the CRB 
is then in fact trivially $+\infty$.

In order to characterize the non regular estimation performance, we 
define  regularized CRBs for which a certain a priori knowledge on the parameter $\theta$,
under the form of  a certain set of equality constraints, is assumed: this set of constraints 
should allow to adjust the parameters that cannot be identified
and then to regularize the estimation problem.
The introduction of a priori information on $\theta$ requires knowledge of
$\theta$ in general, which is not available in practice. However, the point here is to
evaluate the estimation performance (\eg to compare different estimation
algorithms), which implies that we compare
$\hat{\theta}$ to the true $\theta$ which hence needs to be available.
The sample estimation error covariance matrix will furthermore be compared to
the CRB which also depends on $\theta$.

We determine a CRB for estimation under constraints for both cases of regular
and singular  unconstrained estimation problems.
These results are also used in~\cite{carvalho:SBCRB99}, to compare 
blind and semi--blind channel estimation performance under the 
same constraints.

\begin{figure}
\centerline{\includegraphics[width=0.8\columnwidth]{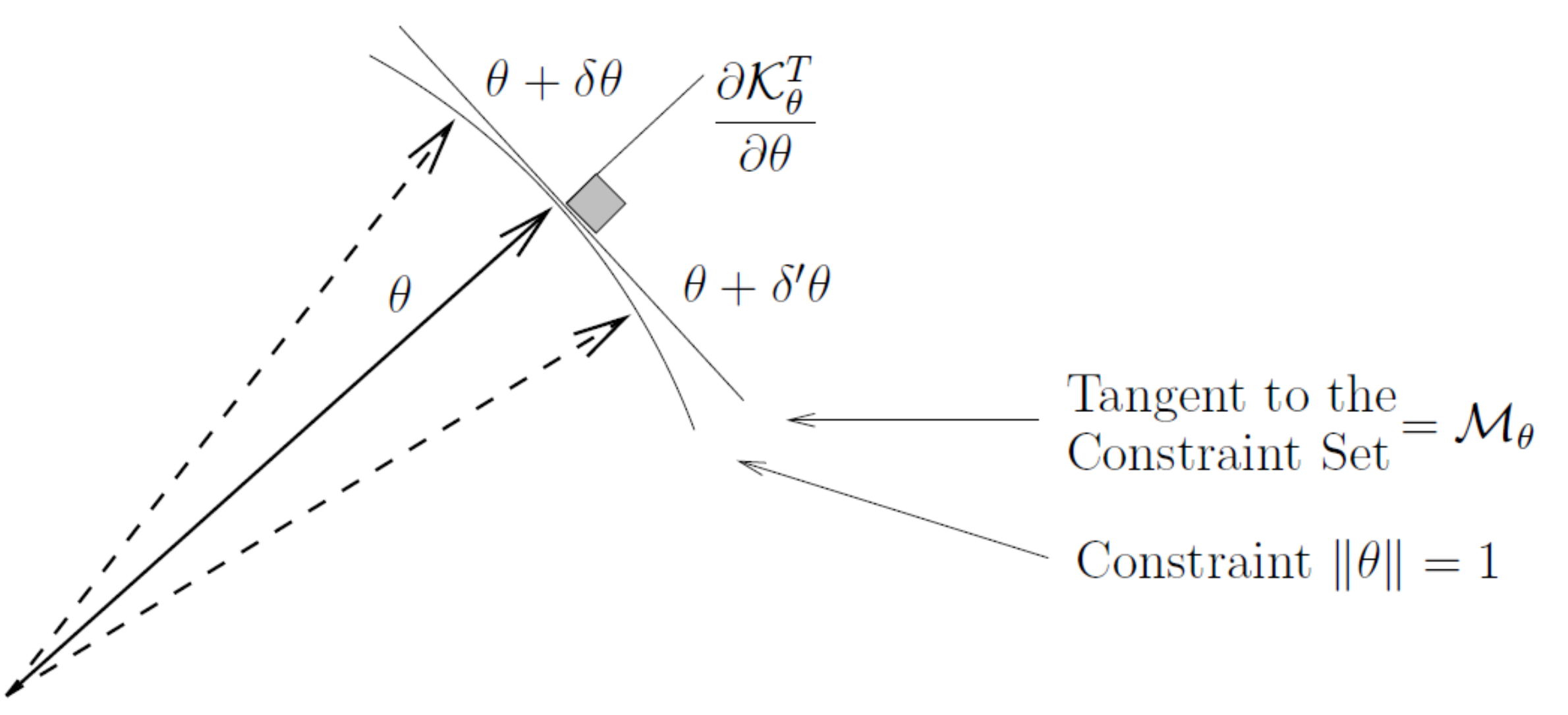}}
\vspace{-2mm}
\caption{Example with constraint $|\theta|=1$.}
\label{C3fig1}
\vspace{-3mm}
\end{figure}

CRBs for parametric estimation under constraints were derived in~\cite{Hero:it90}
in the case where the unconstrained estimation problem is regular.
A simpler derivation of these results was presented in \cite{marzetta:sp93}.
The main ingredient of this simpler derivation was used in
\cite{stoica:spl98} to give an alternative expression for the CRB
in the case where the unconstrained problem is unidentifiable.
We shall succinctly restate these results, which appeared already
in \cite{carvalho:spawc97} for the case of linear constraints.
In these references, and also here, we shall assume that the constraints
are sufficient to regularize the estimation problem, \ie
to render the CRB finite.
So, consider a $k$--fold  constraint of the form:
\beq
\cK_\theta = 0
\label{C2eq21}
\eeq
where $\cK_\theta: \Rr^n \rightarrow \Rr^k$  is continuously differentiable
and $k<n$, $n$ being the number of parameters in the vector $\theta$.
A constrained parameter estimator $\thetah$ is called unbiased if
it satisfies the constraints ($\cK_{\thetah} = 0$) and if the parameter
estimation bias is zero for all parameter values that satisfy the
constraints \cite{marzetta:sp93}. 
The constrained CRB depends on the constraints 
only through the tangents to the constraint set at the true value of $\theta$:
\beq
\cM_{\theta}=
\la Z \in \Rr^n \; ;\;\; 
Z^T\, \dfrac{\partial \cK^T_{\theta}}{\partial \theta} =0 \ra \; .
\label{C2eq28}
\eeq
We note that $\dim(\cM_{\theta})$
can be larger than $n\m k$; the results of \cite{marzetta:sp93} can be
generalized to the case in which the constraints are not independent.
We can introduce a matrix $\cV_\theta$ of full column rank such that
$\cV_{\theta}= \lb \dfrac{\partial \cK^T_{\theta}}{\partial \theta}\rb^{\perp}$,
meaning
\beq
\range\la\cV_\theta\ra = \cM_\theta = \lb \range\la 
\dfrac{\partial \cK^T_{\theta}}{\partial \theta}\ra\rb^\perp \; .
\eeq
The key step now \cite{marzetta:sp93} is that the unbiasedness leads to
a particular correlation between the parameter estimation error and
the loglikelihood gradient restricted to $\cM_\theta$:
\begin{lemma}
Assume the estimator $\thetah$ and the true parameter $\theta$
satisfy the constraints, then unbiasedness implies
\beq
\E\: \cV_\theta^T\dfrac{\partial \ln f(\bmY|\theta)}{\partial \theta}\,
(\thetah - \theta)^T \, =\, \cV_\theta^T \; .
\eeq
\end{lemma}
Using this lemma, the CRB derivation becomes an application
of the following theorem.
\begin{thm}[Cauchy-Schwartz inequality for correlation matrices]
Let $X_1$ and $X_2$ be random vectors with correlation matrices
$R_{ij} = \E X_iX_j^T\,,\; i,j=1,2$. Assume that $R_{11}$ is nonsingular.
Then
\beq
\E\, (X_2-R_{21}R_{11}^{-1}X_1)(X_2-R_{21}R_{11}^{-1}X_1)^T = 
R_{22} - R_{21}R_{11}^{-1}R_{12} \geq 0
\eeq
with equality iff $X_2 = R_{21}R_{11}^{-1} X_1$ in m.s.
\end{thm}
With $X_2 = \thetah - \theta$ and
$X_1 = \cV_\theta^T\dfrac{\partial \ln f(\bmY|\theta)}{\partial \theta}$,
this leads immediately to the following main result.
\begin{thm}[Constrained CRB]
Assume the constrained estimator $\thetah$ to be unbiased
($\thetah$ and $\theta$ satisfy the constraints $\cK_\theta=0$), then 
\beq
C_{\thetat\thetat}\geq CRB_C
 = \cV_\theta \lb \cV_\theta^T \cJ_{\theta\theta} \cV_\theta \rb^{-1} \cV_\theta^T,
\label{C2eq37}
\eeq
with equality iff
\beq
\thetah - \theta \, =\, CRB_C\,
\dfrac{\partial \ln f(\bmY|\theta)}{\partial \theta} \;\; \mbox { in m.s.}
\label{C2eq40}
\eeq
A necessary and sufficient condition for the boundedness of $CRB_C$
is the nonsingularity of $\cV_\theta^T \cJ_{\theta\theta} \cV_\theta$.
\end{thm}

\subsection{Interpretations and Remarks}

The key points to understand how constrained CRBs work are:
\begin{itemize}
\item
the constrained CRB depends on the constraints 
only locally (as the CRB is based on derivatives),
\item 
locally, the constraint set can be linearized.
\end{itemize}
To make things clearer, we distinguish  between the variable $\theta$ and
its true value $\theta^o$. 
Locally, a point $\theta$ belonging to the constraint set can
be approximated as $\theta = \theta^o + \thetat$, where $\thetat$ belongs to
$\cM_{\theta}$, \ie:
\beq
\theta = \theta^o + \cV_{\theta^o} \xi\; .
\label{C2eq29}
\eeq
In figure~\ref{C3fig1}, we show an example with
constraint $|\theta|=1$ (for a complex $\theta$  with $n=1$).

From (\ref{C2eq29}) and applying the chain rule, we get
\beq
\cJ_{\xi\xi} = 
\lb \dfrac{\partial \theta^T}{\partial \xi} \rb
\cJ_{\theta\theta} 
\lb  \dfrac{\partial \theta^T}{\partial \xi} \rb^T
= 
\cV_{\theta^o}^T
\cJ_{\theta\theta} 
\cV_{\theta^o}\; .
\eeq
The estimation of $\xi$ is regular provided that 
$\cV_{\theta^o}^T \cJ_{\theta\theta} \cV_{\theta^o}$ is nonsingular. 
If we now get back to the initial parameter 
$\theta = \theta^o +\cV_{\theta^o}\xi$,
using the CRB for a transformation of parameters~\cite{Kay:book}, we find:
\beq
CRB_{\theta\theta} = 
\lb \dfrac{\partial \theta^T}{\partial \xi} \rb^T
\cJ_{\xi\xi}^{-1}
\lb \dfrac{\partial \theta^T}{\partial \xi} \rb 
= \cV_{\theta^o}
\lb \cV_{\theta^o}^T \cJ_{\theta\theta}  \cV_{\theta^o} \rb^{-1}
 \cV_{\theta^o}^T =CRB_C\; .
\label{C2eq36}
\eeq
We see that the constrained CRB can be interpreted in terms
of regular estimation: 
$\cV_{\theta}^T \cJ_{\theta\theta}\cV_{\theta}$ 
can be considered as the FIM for a minimal parameterization $\xi$
of $\theta$, and the
results of equivalence mentioned in section~\ref{section41} 
between FIM and local identifiability could also be applied here. 

The $CRB_C$ is independent of the choice of $\cV_\theta$ and can in fact
also be written as:
\beq
CRB_C = A_\theta \lb A_\theta^T \cJ_{\theta\theta} A_\theta \rb^{+} A_\theta^T
\label{C2eq38}
\eeq
where
$A_\theta$ is a $n \times{}q$ matrix, $q \geq \dim(\cM_{\theta})$, 
such that $\cM_{\theta} = \range\la A_{\theta}\ra$. In particular,
denoting $\dfrac{\partial \cK^T_{\theta}}{\partial \theta} = \cK'_{\theta}$,
we can take $A_{\theta}= P^\perp_{\cK'_{\theta}}=P_{\cV_{\theta}} $
(where $P_X^{\perp} = I-P_X = I-X(X^HX)^+X^H$) and obtain
\beq
CRB_C=
P_{\cV_{\theta}} 
 \lb P_{\cV_{\theta}} \cJ_{\theta\theta}
P_{\cV_{\theta}} \rb^{+} P_{\cV_{\theta}} =
 \lb P_{\cV_{\theta}} \cJ_{\theta\theta} P_{\cV_{\theta}} \rb^{+} \; .
\label{C2eq39}
\eeq

There is a direct correspondence between the number of FIM 
singularities and the number of constraints necessary to have a finite constrained 
CRB, which is also the number of constraints necessary to have local identifiability.
\begin{thm}
For the constrained CRB to be defined, it is necessary and sufficient 
to fulfill the following conditions.
\begin{itemize}
\item
The number of independent constraints should be at least equal to $n-r$ 
($r=\rank(\cJ_{\theta\theta})$).
\item
At least $n-r$ independent columns of $\dfrac{\partial \cK^T_{\theta}}{\partial
\theta}$ should not be orthogonal to the null space of $\cJ_{\theta\theta}$.
\end{itemize}
\end{thm}

A constraint of the form $\cK_\theta=0$ has only a local effect  and 
can be locally linearized
\begin{thm} The constrained CRB (\ref{C2eq37})
 can also  be interpreted as the CRB under the linear constraint:
\beq
\theta^T\, \left. \dfrac{\partial \cK_\theta^T}{\partial
  \theta}\right|_{\theta=\theta^o} \, =\, \theta^{o\, T}
\left. \dfrac{\partial \cK_\theta^T}{\partial \theta}\right|_{\theta=\theta^o} 
\label{C2eq41}
\eeq
which means that the components of $\theta$ 
in $\mbox{span}\la\left. \dfrac{\partial \cK_\theta^T}{\partial \theta} \right|_{\theta=\theta^o}\ra$
are known.
\end{thm}

%------------------------------------------
\subsection{Minimal constrained CRB}
%------------------------------------------
\label{section53}

\subsubsection{Constraint on the Global Parameter}

We assume here that $\cJ_{\theta \theta}$ is singular.
When $\range\la\cV_\theta\ra=\range\la \cJ_{\theta\theta}\ra$ 
and since $\cV_\theta$ has full column rank,
$\cV_\theta^T \cJ_{\theta\theta} \cV_\theta$ is regular (minimal number
of independent constraints in this case) and the 
constrained CRB is:
\beq
CRB_C = \cJ_{\theta\theta}^+ \; .
\label{C2eq42}
\eeq
This is a particular constrained CRB:  
we prove in \ref{C3app1} that, 
among all sets of a minimal number of independent constraints, $CRB_C = \cJ_{\theta\theta}^+$
yields the lowest value for $\trace \la CRB_C \ra$.
This means that if we want to introduce a priori information in the form
of independent constraints, enough to regularize the estimation problem,
but not more (minimal number), then all the constraints should concentrate on the unidentifiable
part of the parameters
($\range\la\dfrac{\partial \cK_\theta^T}{\partial \theta}\ra=\mbox{null}\la\cJ_{\theta\theta}\ra$)
to minimize $\trace \la CRB_C \ra$.

\subsubsection{Constraint on a Parameter Subset}
\label{section532}

Consider the case of the joint estimation of two parameter vectors $\theta_1$ and
$\theta_2$ of length $n_1$ and $n_2$ ($n_1+n_2=n$). 
We are interested in the estimation of $\theta_1$ with  $\theta_2$ being 
nuisance parameters.
The overall parameter vector is 
$\theta=\lsb \theta_1^T \;\; \theta_2^T\rsb^T$.
\beq
\cJ_{\theta\theta}= 
\lsb 
\begin{array}{cc}
\cJ_{\theta_1\theta_1}   &  \cJ_{\theta_1\theta_2} \\
\cJ_{\theta_2\theta_1}   &  \cJ_{\theta_2\theta_2}
\end{array}
\rsb \; . 
\label{C2eq43}
\eeq
We consider the case in which $\cJ_{\theta\theta}$ is singular
but $\cJ_{\theta_2\theta_2}$ is regular.
%The question is which constraint minimizes
%$\trace \la CRB \ra$.
To regularize the estimation problem, we consider the introduction of
(independent) constraints on $\theta_1$ only: $\cK_{\theta_1}=0$, 
$\cK_{\theta_1}: \Rr^{n_1} \rightarrow \Rr^{k_1}$.
\beq
\dfrac{\partial \cK^T_{\theta_1}}{\partial \theta} = 
\lsb
\begin{array}{l}
\dfrac{\partial \cK^T_{\theta_1}}{\partial \theta_1} \\
0_{n_2,k_1}
\end{array}
\rsb
\label{C2eq44}
\eeq
(assumed full column rank).
Let $\cV_{\theta_1}=\lb \dfrac{\partial \cK_{\theta_1}}{\partial \theta_1}\rb^{\perp}$
be a $n_1\times{}(n_1-k_1)$ matrix of full column rank.
We can choose
\beq
\cV_\theta=
\lsb
\begin{array}{cc}
\cV_{\theta_1}   &  0_{n_1,n_2}  \\
0_{n_2,n_1}    &  I_{n_2,n_2}
\end{array}
\rsb\; .
\label{C2eq45}
\eeq
The constrained CRB for $\theta$ is:
\beq
CRB_C=\cV_\theta \lb \cV_\theta^H \cJ_{\theta\theta} \cV_\theta \rb^{-1} \cV_\theta^H= 
\cV_\theta
\lsb
\begin{array}{cc}
\cV_{\theta_1}^H \cJ_{\theta_1\theta_1}\cV_{\theta_1}  &  \cV_{\theta_1}^H \cJ_{\theta_1\theta_2} \\
\cJ_{\theta_2\theta_1} \cV_{\theta_1}  &  \cJ_{\theta_2\theta_2}
\end{array}
\rsb^{-1}\cV_\theta^H
\label{C2eq46}
\eeq
and the constrained CRB for $\theta_1$ separately is:
\beq
CRB_{C,\theta_1}=
\cV_{\theta_1} \lb  \cV_{\theta_1}^H \lsb \cJ_{\theta_1\theta_1} - 
\cJ_{\theta_1\theta_2} \cJ_{\theta_2\theta_2}^{-1} \cJ_{\theta_2\theta_1}\rsb \cV_{\theta_1}  \rb^{-1} \cV_{\theta_1}^H
= \cV_{\theta_1} \lb  \cV_{\theta_1}^H \cJ_{\theta_1\theta_1}(\theta) \cV_{\theta_1}  \rb^{-1} \cV_{\theta_1}^H
\label{C2eq47}
\eeq
where we introduced the notation $\cJ_{\theta_1\theta_1}(\theta)$ for
$\cJ_{\theta_1\theta_1} - \cJ_{\theta_1\theta_2} \cJ_{\theta_2\theta_2}^{-1}
\cJ_{\theta_2\theta_1}$. This notation will be reused below.
Note that $\cJ_{\theta_1\theta_1}^{-1}(\theta)$ would be the unconstrained CRB for
$\theta_1$ if $\cJ_{\theta\theta}$ were regular.
Note that with $\cJ_{\theta_2\theta_2}$ being regular,
$\cJ_{\theta_1\theta_1}(\theta)$ has the same number of singularities
as $\cJ_{\theta\theta}$ in the singular case.
Now assume that the constraints are such that
$\range\la\cV_{\theta_1}\ra= \range\la \cJ_{\theta_1\theta_1}(\theta)\ra$. 
Then it can be proven that such constraints give the minimal constrained CRB
for $\theta_1$, 
\beq
CRB_{C,\theta_1} = \cJ_{\theta_1\theta_1}^+ (\theta) 
\label{C2eq48}
\eeq 
over all sets of a minimal number of independent constraints on 
$\theta_1$ only.

%\paragraph{Remark} When the FIM is singular, Hero in~\cite{Hero:sp96} claims
%that the pseudo-inverse of the FIM would be a lower estimation bound, 
%without mentioning any assumed constraints. 
%This result is false. Indeed, the fact that the FIM is singular comes  from 
%the fact that all the parameters cannot be estimated and that constraints
%must be used. ((He ignores these constraints and introduced the mean of the 
%estimator, which is not defined without constraints.))
%
%The constraint corresponding to $K_\theta$ is a ``minimal'' constraint.
%Indeed, let $K_\theta'$ corresponding to another constraint which columns does 
%not contain  the entire null space of $J_{\theta\theta}$,  ${K'_\theta}^T
%J_{\theta\theta} K'_\theta$ 
%is not regular, the constraints are not sufficient to allow estimation of all
%the  parameters. 
%Now if the columns of $K_\theta'$ span the entire null space of
%$J_{\theta\theta}$ as well as a certain additional space, then:
%\beq
%K'_\theta \lb {K'_\theta}^T J_{\theta\theta} K'_\theta \rb^{-1} {K'_\theta}^T  
%\leq J_{\theta\theta}^+
%\label{C2eq37}
%\eeq
%
%According to the previous section, it can be interpreted
%as the CRB under the linear constraint:
%${V_2^o}^H h = {V_2^o}^H h^o$
%where the columns of $V_2^o$ form a basis of the null space of  $J^o_{\theta\theta}$.

%%%%%%%%%%%%%%%%%%%%%%%%%%%%%%%%%%%%%%%%%%%%%%%%%%%%%%%%%%%%%%%%%%%%%%%%%%%%%%%%%%%%%%%%%%%%
%%%%%%%%%%%%%%%%%%%%%%%%%%%%%%%%%%%%%%%%%%%%%%%%%%%%%%%%%%%%%%%%%%%%%%%%%%%%%%%%%%%%%%%%%%%%

\section{CRBs for Blind FIR Multichannel Estimation}

These results are now applied to blind FIR multichannel estimation.
Two models are presented here: the deterministic model which considers the 
symbols as deterministic quantities and the Gaussian model which considers
them as Gaussian random variables~\cite{carvalho:Ident99}.
We first present the multichannel model, which is fundamental in 
blind channel estimation (from second--order statistics).

%%%%%%%%%%%%%%%%%%%%%%%%%%%%%%%%%%%%%%%%%%%%%%
\subsection{The Multichannel Model}
%%%%%%%%%%%%%%%%%%%%%%%%%%%%%%%%%%%%%%%%%%%%%%

We consider a single-user multichannel model: see~\cite{carvalho:Ident99} for  
more details on this model.
Let $a(k)$be a sequence of symbols  received through $m$ channels 
of length N and with impulse response $\bmh(i)$:
\beq
\bmy(k) =\sum_{i=0}^{N-1} \bmh(i) a(k\m i)+\bmv(k)
\label{C2eq49},
\eeq
$\bmy(k)=\lsb y_{1}(k) \; \cdots \; y_{m}(k)\rsb^T$, 
and similarly for $\bmh(k)$ and $\bmv(k)$.
$\bmv(k)$ is an additive independent white Gaussian noise with
$r_{\bmv\bmv}(k\m i) = \,\E\, \bmv(k)\bmv(i)^{\T} =
\sigma_v^2 I_m \,\delta_{ki}$ and when $\bmv(k)$ is complex 
$\E\, \bmv(k)\bmv^{T}(i) = 0$ (circular noise).

Throughout the paper, we will distinguish between complex and real symbols.
For real symbols, it will be advantageous to consider separately the real and
imaginary parts of the channel and received signal:
\beq
\lsb\begin{array}{c} \re(y_l(k)) \\ \im(y_l(k)) \end{array}\rsb
= \dsum_{i=0}^{N-1} 
\lsb\begin{array}{c} \re(h_l(i)) \\ \im(h_l(i)) \end{array}\rsb
\, a(k\m i) + 
\lsb\begin{array}{c} \re(v_l(k)) \\ \im(v_l(k)) \end{array}\rsb
\;, \;\; l=1,\ldots, n
\label{C2eq50}
\eeq
where $n$ now denotes the product of the oversampling factor and the number
of sensors. The vector signals now become
$\bmy(k) = \lsb \re(y_1(k))\; \im(y_1(k)) \cdots \re(y_n(k))\; \im(y_n(k))
\rsb^T$ and similarly for $\bmh(k)$ and $\bmv(k)$.
This leads to a representation similar to (\ref{C2eq49}).
The number of channels gets doubled, which has for advantage to 
increase diversity. Note that the monochannel  case does not exist 
in the real case, unless transmission is performed in baseband.

Assume we receive $M$ samples, concatenated in the vector $\bmY_M(k)$:
\beq
\bmY_M(k)  =  \cT_M(h)\, A_{M+N-1}(k) + \bmV_M(k)
\label{C2eq51}
\eeq
where $\bmY_M(k) = [\bmy^{T}(k) \cdots \bmy^{T}(k\m M\p 1)]^{T}$,
and $\bmV_M(k)$ and $A_M(k)$ are similarly defined.
%=\lsb a^H(k) \cdots a^H(k\m M\m N\p 2)\rsb^{H}.
$\cT_M(h)$ is a block Toeplitz matrix with $M$ block rows and $\lsb \bmH \;\;
0_{m\times{}(M-1)}\rsb$ as first block row: 
\beq 
\bmH=\lsb \bmh(0) \; \cdots \; \bmh(N-1) \rsb
\mbox{ and } 
h=\lsb \bmh^T(0)  \cdots \bmh^T(N\m 1)\rsb^T \; .
\label{C2eq52}
\eeq
We furthermore denote $\bfH(z) = \sum_{i=0}^{N-1}\bmh(i)z^{-i}=
[\rmH_1(z) \cdots \rmH_m(z)]^{T}$, the SIMO channel transfer
function.
We shall simplify the notation in (\ref{C2eq51}) with $k=M\m 1$ to:
\beq
\vspace{0mm}
\bmY \; = \; \cT(h)\, A + \bmV \; .
\label{C2eq53}
\eeq

\paragraph{Commutativity of Convolution}
We will need the convolution commutativity relationship:
\beq
\cT(h)  A = \cA h
\label{C2eq54}
\eeq
where: $\cA  = A' \otimes I_m$,
\beq
A' = 
\lsb
\begin{array}{cccc}
a(-N\p 1)     &  a(-N\p 2)  &  \cdots  &  a(0)\\
a(-N\p2)   &  \adots  &  \adots  & \vdots  \\
\vdots   &  \adots  &  \adots & \vdots  \\
a(M\m N \p 2) &  \cdots  &  \cdots  &  a(M\m 1)
\end{array}
\rsb \; .
\label{C2eq55}
\eeq

\paragraph{Irreducible and Reducible Channels}
A channel is said irreducible if its subchannels $\rmH_i(z)$
have no zeros in common, and reducible otherwise. 
A reducible channel can be decomposed as:
\beq
\bfH(z)=\bfH_I(z) \rmH_c(z),  
\label{C2eq56}
\eeq
where $\bfH_I(z)$ of length $N_I$ is
irreducible and $\rmH_c(z)$  of length $N_c=N-N_I+1$ is a monochannel.
%for which is assume $\rmH_c(\infty)=1$.

%%%%%%%%%%%%%%%%%%%%%%%%%%%%%%%%%%%%%%%%%%%%%%
\subsection{Deterministic Model}
%%%%%%%%%%%%%%%%%%%%%%%%%%%%%%%%%%%%%%%%%%%%%%

The deterministic  model considers the joint estimation
of the unknown input symbols $A$ and  the channel coefficients $h$.
The parameter vector is  $\theta=\lsb A^T \;  h^T \rsb^T$.
It can indeed 
be shown that the estimation of $A$ and $h$ is decoupled from that of
$\sigma_v^2$, at the FIM level (and also from an estimation point of view).
So we shall omit $\sv$ in $\theta$. Identifiability of $(A,h)$ occurs from
$m_Y(\theta)=\bmX= \cT(h) A$, the signal part of $\bmY$, 
  whereas identifiability of $\sv$ occurs from
$C_{YY}(\theta)=\sv I$. Blind identifiability in the deterministic model requires
the channel to be irreducible and the burst length and the number of input 
excitation modes to satisfy certain minimum requirements~\cite{carvalho:Ident99}.

%------------------------------------------
\subsubsection{FIMs}
%------------------------------------------

\paragraph{Circular Complex Input Constellation}

As $\bmY$ is circular, 
we can work with  the complex probability density function of the Gaussian
random variable $\bmY \sim \cN(\cT(h)A,\sv I )$. 

As $J_{\theta\theta^*}=0$, the complex FIM $J_{\theta\theta}$ is equivalent
to the real one $\cJ_{\theta_R\theta_R}$ and is equal to:
\beq
J_{\theta\theta}=
\dfrac{1}{\sigma_v^2}
\lb \dfrac{\partial \bmX^H}{\partial {\theta^*}}\rb 
\lb \dfrac{\partial \bmX^H}{\partial {\theta^*}}\rb^H
=\dfrac{1}{\sigma^2_v}
\lsb
\begin{array}{c}
\cT^H(h) \\ \cA^H
\end{array} 
\rsb
\lsb
\begin{array}{cc}
\cT(h) & \cA 
\end{array} 
\rsb
\label{C2eq57}
\eeq
because  $\dfrac{\partial \bmX^H}{\partial A^*} = \cT^H(h)
\mbox{ and }
\dfrac{\partial \bmX^H}{\partial {h^*}}=\cA^H.
$ 

\paragraph{Real Symbol Constellation}

The FIM is the same as in (\ref{C2eq57}).
This equality of the expressions will allow us to treat the
complex and real cases simultaneously.\\

%------------------------------------------
\subsubsection{Singularities of the FIMs}
%------------------------------------------

Under the blind deterministic identifiability conditions, 
$(h,A)$ are identifiable up to a scale factor: 
indeed $\cT(h)A = \cT({h}/{\alpha})\alpha {A}$. 
This results in one (complex) singularity of the complex FIM (see theorem below).
We examine here the singularities in that case.  
The singularities of the FIM can be considered at the level of $\theta$ (joint
estimation of $A$ and $h$) or at the level of $h$ (estimation of $h$
with $A$ considered as nuisance parameter).

%\begin{itemize}
%\item
\paragraph{Singularities of $J_{\theta\theta}$.}
 $J_{\theta\theta}=\dfrac{1}{\sigma_v^2}
\lsb \cT(h)\;\; \cA\rsb^{\T} \lsb \cT(h)\;\; \cA\rsb$ admits as 
null vector: $\theta_s=\lsb \m A^T \;\; h^T \rsb^T$. Indeed, 
$\lsb \cT(h)\;\; \cA\rsb \lsb \m A^T \;\; h^T \rsb^T = \m \cT(h)A + \cA h =0$,
by exploiting (\ref{C2eq54}). When 
$\cT(h)$  and $\cA$ have full column rank, the nullity of 
$J_{\theta\theta}$ is the dimension of the intersection of the 
column spaces of $\cT(h)$ and $\cA$, which is one.

\paragraph{Singularities of $J_{hh}(\theta)\stackrel{\triangle}{=} \dfrac{1}{\sv}\cA^H
  P^\perp_{\cT(h)} \cA$.} If $J_{hh}(\theta)$ were regular, its inverse would be 
the CRB for $h$. However, $J_{hh}(\theta)$ is singular.
Indeed, assume that  $h'$ is a singular vector of $J_{hh} (\theta)$: $\cA^H
P^\perp_{\cT(h)} \cA h'=0$.
Then, as  $\cA$ has full column rank, this means that $\cA h' \in \range\{
\cT(h)\}$:  there exists an $A'$ such that $\cA h' = \cT(h') A = \cT(h)A'$.
This last relation is satisfied for $(h',A')=(h,A)$.
Hence, $J_{hh}(\theta)$ has one singularity with
$h$ as its singular vector
($\cA^H P^\perp_{\cT(h)} \cA h = \cA^H P^\perp_{\cT(h)} \cT(h) A = 0$)
and the singularity of $J_{hh}(\theta)$ is due to the same mechanism
that leads to the singularity of the global FIM $J_{\theta\theta}$.\\
In the complex case, 
$\cJ_{\theta_R\theta_R}$ will have 2 singularities spanned by:
\beq
h_{S_1}=
\lsb
\begin{array}{r}
\re(h)\\
\im(h)
\end{array}
\rsb = h_R
\hspace{5mm} \mbox{ and } \hspace{5mm}
h_{S_2}=
\lsb
\begin{array}{r}
-\im(h)\\
\re(h)
\end{array}
\rsb
=
\lsb
\begin{array}{r}
\re(j h)\\
\im(j h)
\end{array}
\rsb \; .
\label{C2eq58}
\eeq
The first null vector can be interpreted as corresponding  
to the ambiguity in the norm of the channel
and the second one to the ambiguity in the phase factor.

\subsubsection{Equivalence between FIM Regularity and  Local Identifiability}
\label{secdet}

The link between blind identifiability and FIM singularities in the 
specific case of the blind deterministic model was
already studied in~\cite{Hua:sp96,HuaWax:sp96}:
\begin{thm}
For $M\geq 2(N-1)$, or $M \geq N$ for 2 subchannels ($m=2$), 
a channel is blindly identifiable up to a scale factor if and only if 
the complex FIM $J_{\theta\theta}$ has exactly one singularity.
\end{thm}
{\em Proof: }see~\cite{HuaWax:sp96}.

\hfill $\square$

\vspace{7mm}
In general, there is a correspondence between local identifiability and
FIM regularity.
\begin{thm}
A channel is locally blindly identifiable up to a scale factor if and only if 
the complex FIM $J_{\theta\theta}$ has exactly one singularity.
\label{deteq}
\end{thm}
{\em Proof:}
Assume that the FIM has a singular vector $\theta'=\lsb {h'}^T\;\; {A'}^T\rsb^T$
different
from $\lsb {h}^T\;\; -{A}^T\rsb^T$:
\beq
\cT(h) A' +\cT(h')A = 0 \; .
%\;\; \Leftrightarrow \;\;
%\cT(h) \eps A' +\cT(\eps h')A = 0
\label{C2eq59}
\eeq
Then for $\eps >0$ arbitrarily small:
\beq
\begin{array}{rcl}
m_Y(\theta+\eps \theta') - m_Y(\theta) & = & 
\cT(h+\eps h') \lsb A + \eps A'\rsb - \cT(h) A \\
& = & \eps \lsb \cT(h) A' + \cT(h')A \rsb + O(\eps^2) \\
& = &  O(\eps^2) 
\end{array}
\label{C2eq60}
\eeq
which implies that $\theta$ is not locally blindly identifiable.

Now assume that $\theta$ is not locally blindly identifiable, then one can find 
$\Delta h$ and $\Delta A$, where $\|\Delta h\|$ and $\|\Delta A\|$ are
arbitrarily small, and not simultaneously colinear with  $h$ and $A$ resp.\ verifying
$\cT(h) A = \cT(h+\Delta h) (A+\Delta A)$.  Then:
\beq
\begin{array}{rcl}
\cT(h+\Delta h) (A+\Delta h) - \cT(h)A  & = & \cA \Delta h + \cT(h)\Delta A
 +O(\|\Delta h\|\|\Delta A\|)  \\
& = & 0 \; .
\end{array}
\label{C2eq61}
\eeq
This implies that  $\lsb \Delta {h}^T\;\; \Delta {A}^T\rsb^T$ is 
a singular vector of the FIM,
non colinear to $\lsb {h}^T\;\; \m{A}^T\rsb^T$.

\hfill $\square$

Using a similar derivation, we can also show the equivalence between the 
regularity of  $\cV^H_\theta J_{\theta\theta}\cV_\theta$  
and local identifiability  under constraint $\cK_\theta$:
\begin{thm}
The estimation  problem under constraint $\cK_\theta$ is locally identifiable 
if and only if the regularized FIM $\cV^H_\theta J_{\theta\theta}\cV_\theta$ 
(with definitions of section~\ref{sectionconst}) is regular.
\end{thm}
The same theorem will hold for the Gaussian model in section~\ref{secgaus} also
but will not be restated there.

\subsubsection{Regularized Blind CRB}
\label{regCRB}

To define a  regularized blind CRB, we assume some a priori knowledge.
Different forms of a priori knowledge are possible.
A technique often used consists in assuming a coefficient of 
the channel to be known. 
This is however not robust as performance depends heavily
on the choice of this known coefficient (which can be arbitrarily small).  
The proposed  regularized CRB, the Moore--Penrose pseudo--inverse of
the FIM,  appears less arbitrary and is directly related to the blind scale
factor ambiguity. 

Blind methods commonly consider the quadratic constraint: $h^H h = 1$
(see~\cite{stoica:spl98}). 
This constraint does not render the problem identifiable: it leaves a sign 
ambiguity when $h$ is real and a continuous phase ambiguity when $h$
is complex. 
In the former case, the computation of mean squared error (MSE)
%(\ie $\left\|\hh - \dfrac{h^o}{\|h^o\|}\right\|$)
assumes the right sign 
(the right sign could be taken as the sign giving the smallest error).
In the complex case however, which phase factor should be chosen?
A frequent choice consists in imposing one element of $h$ to be real 
and positive; 
again the resulting CRB  depends on the choice of this element. 

The blind regularized CRB proposed here is computed under the following constraints:
\begin{enumerate}[(1)]
  \item A quadratic constraint: 
\beq
h^H h= {h^o}^H h^o
\label{C2eq62}
\eeq
(equivalent to the usual constraint $h^H h= 1$) which allows to adjust the norm of the channel.
%In the real case, this constraint is sufficient as the remaining sign ambiguity 
%will not lead to a FIM singularity: the constrained CRB assumes the right sign.
    \item In the complex case, an additional constraint is necessary to adjust 
the  phase factor:
\beq
h_{S_2}^{o\, T} h_R =  h_{S_2}^{o\, T} h_R^o =0 \; .
\label{C2eq63}
\eeq
\end{enumerate}
In both real and complex cases, these constraints leave a sign ambiguity which
does not lead to FIM singularity. For MSE evaluation, the ambiguity can be
resolved by requiring $h^{o\, T}h >0$. 

\begin{result}
Under constraint (\ref{C2eq62}) (and (\ref{C2eq63}) for the complex case) the
CRB for $h$ is the 
Moore-Penrose pseudo-inverse of $J_{hh}(\theta)$:
\beq
CRB_{C,h} = J^+_{hh}(\theta)= \sv\lb \cA^H P^\perp_{\cT(h)} \cA \rb^+ \; .
\label{C2eq64}
\eeq
\end{result}

\noindent
{\em Proof:}
%%We prove here that the CRB for $h$ under constraint (\ref{}) (and (\ref{})) is $J^+_{hh}(\theta)$.
Following the notations of section~\ref{sectionconst}, the constraint is:
\beq
\cK_{h_R} = 
\lsb
\begin{array}{c}
h_R^T h_R - {h^o}_R^T h^o_R \\
h_{S_2}^{o\, T} h_R
\end{array}
\rsb 
=0
\label{C2eq65}
\eeq
leading to 
\beq   
\dfrac{\partial \cK^T_{h_R}}{\partial h_R} = 
\lsb 
2 h_R^o \;\;
h_{S_2}^{o}
\rsb.
\label{C2eq66}
\eeq
As $h_R$ and $h_{s_2}$ are the singular vectors of $\cJ_{h_R h_R}(\theta)$ (which
corresponds to the previously defined complex $J_{hh}(\theta)$), 
the orthogonal complement of $\range\la\dfrac{\partial
\cK^T_{\theta_R}}{\partial h_R}\ra$ equals $\range\la\cJ_{h_R h_R}(\theta)\ra$.
According to section~\ref{section532}, the CRB under constraint (\ref{C2eq65}) is:
\beq
CRB_{C,h_R} = \cJ_{h_R h_R}^+(\theta)
\label{C2eq67}
\eeq
and the corresponding complex contrained CRB can be shown to be:
\beq
CRB_{C,h} = J_{h h}^+(\theta)
\label{C2eq68}
\eeq

\hfill $\square$

The choice of the first constraint is not only motivated by its common use.
As mentioned in section~\ref{section53}, this constraint 
(possibly combined with  the linear constraint on the phase) 
leads to the minimal constrained CRB.

\begin{figure}
\centerline{\includegraphics[width=0.99\columnwidth]{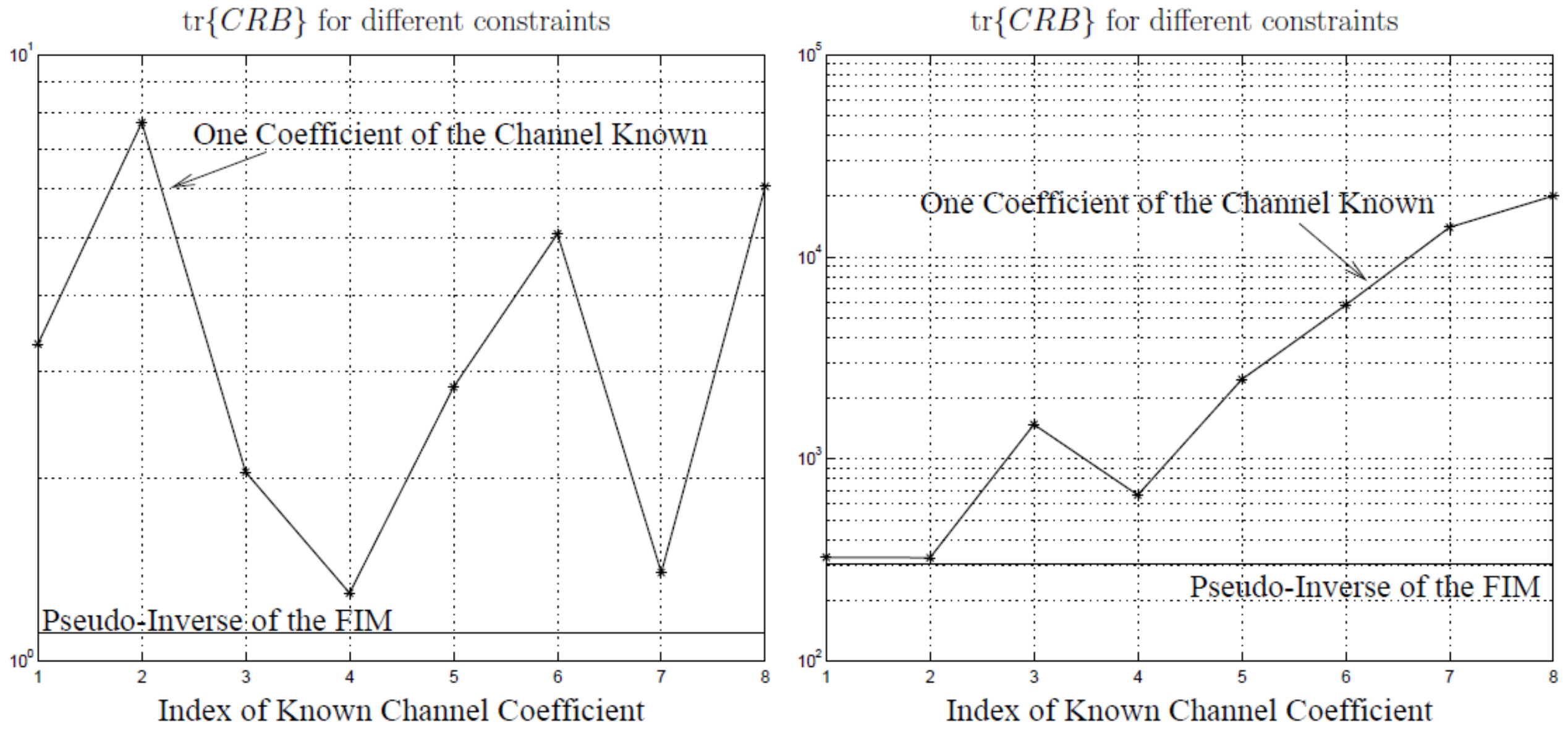}}
\vspace{-4mm}
\caption{Comparison between CRBs with different a priori knowledge. The
  coefficients designate the coefficients of the vector $h$ for the random 
channel $\bmH_1$ (left) 
and the decreasing channel impulse response $\bmH_2$ (right).}
\label{C3fig2}
\vspace{-2mm}
\end{figure}

In Figure~\ref{C3fig2}, we illustrate the importance of the choice for a
constraint by comparing the proposed CRB $\trace\{J^+_{hh}(\theta)\}$ 
to a constrained CRB for which one of the channel coefficients
(of varying position) is 
assumed known. Two channels are tested: a randomly chosen channel and
a channel with slowly decreasing coefficients:
\beq
\bmH_1=
\lsb
\begin{array}{rrrr}
    0.9477 &  -1.1156 &   1.1748  &  1.6455\\
   -0.5257 &  -1.5923 &   0.4851  & -0.4542\\
\end{array}
\rsb
\label{C2eq69}
\eeq
\beq
\bmH_2=
\lsb
\begin{array}{rrrr}
  1.0000  &  0.5000  & -0.1500  &   0.0695\\  
  1.5000  &  -0.9500 &   0.3050  &  0.0550
\end{array}
\rsb
\label{C2eq70}
\eeq
One observes that when the channel coefficient which is assumed known
is small, the corresponding $CRB_C$ can get quite large (arbitrarily large
as the coefficient magnitude shrinks).

\paragraph{Some Equivalent Constraints}

Another choice for the constraint, which leads to the same
$\range\la\dfrac{\partial \cK^T_{h_R}}{\partial h_R}\ra$ is the following 
linear constraint:
\beq
h^{o\,H} h = h^{o\,H} h^o \; .
\eeq
This constraint, which leaves no sign ambiguity, corresponds to forcing the
components of $h$ in the nullspace of $J_{hh}$ to their true values.

Often, $h$ is estimated under a unit  norm constraint $\|\hh\|=1$, and the scale factor 
is adjusted in different ways. The following adjustments lead to the 
same minimal CRB. 
\begin{itemize}
\item
The norm of the channel is adjusted so that 
$\|\hh\|=\|h^o\|$ and the phase using
    the  phase constraint (\ref{C2eq63}). We denote the resulting
estimate $\hhh_{NO}$. 
\item
The scale factor is adjusted in the least--square sense \cite{Bapat:thesis},\cite{morgan:spl98}:
   $\min_\alpha\|h^o-\alpha \hh\|^2$. To be more precise, in this case
 the trace of the corresponding constrained CRB is 
$\trace\la CRB_C \ra$ of equation (\ref{C2eq64}).

\noindent
{\em Proof: } The solution of the least-squares problem is 
$\hhh_{LS}=\hat{\alpha}\hh = P_{\hh} h^o$.  
Then, 
$\Delta \hhh = P_{\hh} h^o - h^o = - P^\perp_{\hh} h^o$;
$C_{\Delta \hhh\Delta \hhh}=\,\E\, P^\perp_{\hh} h^o h^{o\,H} P^\perp_{\hh}$.
\[
\begin{array}{rcl}
\trace\la C_{\Delta \hhh\Delta \hhh} \ra & = &
\trace\la\E\, P^\perp_{\hh} P_{h^o}\ra \|h^o\|^2 =
\trace\la \E\, P^\perp_{h^o}\hh \hh^H \ra \|h^o\|^2\\  
& = & \trace\la\E\, P^\perp_{h^o}(\hh\|h^o\|)(\hh\|h^o\|)^H P^\perp_{h^o}\ra  \\
& = & \trace\la\E\, P^\perp_{h^o}(\Delta\hhh_{NO})(\Delta\hhh_{NO})^H P^\perp_{h^o}\ra
\; = \;  \trace\la P^\perp_{h^o} C_{\Delta \hhh_{NO}\Delta \hhh_{NO}}P^\perp_{h^o} \ra\\
 &\geq &
\trace\la P^\perp_{h^o} CRB_{C,h}  P^\perp_{h^o} \ra = \trace\la CRB_{C,h}\ra
\end{array}
\]

\hfill $\square$
\end{itemize}

Another way to adjust the scale factor consists of adjusting
$\alpha$ by the following  linear constraint 
$h^{o\,H} \hhh_{LIN} = h^{o\,H} \alpha \hh = h^{o\,H} h^o$, leading to the
following channel estimate:
\beq
\hhh_{LIN}=\dfrac{\hh {h^o}^H}{{h^o}^H \hh} h^o\; .
\eeq
When the estimation of $h$ is consistent, then, asymptotically, the CRB for
this constrained channel estimate is the same $CRB_{C,h}$ of (\ref{C2eq64}).

\begin{figure}
\centerline{\includegraphics[width=0.4\columnwidth]{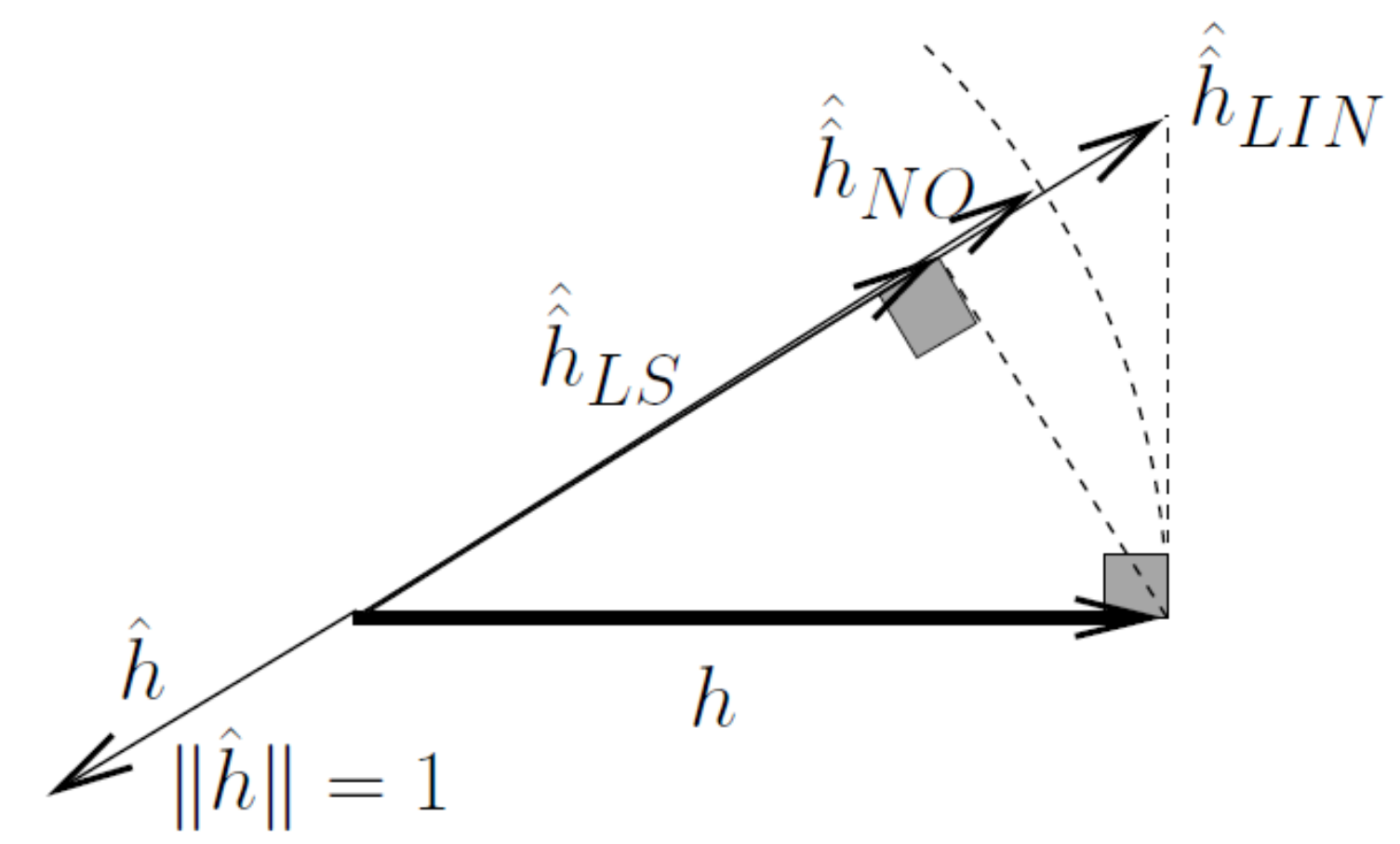}}
\vspace{-4mm}
\caption{Deterministic case: asymptotically equivalent constraints.}
\vspace{-3mm}
\label{C3fig3}
\end{figure}

In figure~\ref{C3fig3}, we show the solutions $\hhh_{NO}$, $\hhh_{LS}$, $\hhh_{LIN}$
for a real channel  of length $N=1$ and with 2 subchannels.

\subsubsection{Reducible Channel Case}

In this case, $\bmH(z)=\bmH_I(z) \rmH_c(z)$ where $\rmH_c(z)$ is a monic
(first coefficient equal to 1) polynomial in $z^{-1}$. In the time domain,
this becomes $h = T_c\, h_I$ where $T_c = \cT^T_{N_I}(h_c)\otimes I_m$.
This decomposition leads us to introduce $A_I= \cT(h_c)A$, the input
signal filtered by $\rmH_c(z)$ and we can write the noise-free received
signal
in the following ways
\beq
\begin{array}{rcl}
\bmX = \cT(h)\, A & = &
\cT(h_I)\, \cT(h_c)\, A = \cT(h_I)\, A_I = \cA_I\, h_I \\
 & = & \cA\, h = \cA T_c \, h_I = \cA_I\, h_I
\end{array}
\label{eq6.2.5.1}\eeq
where $\cA_I=\cA T_c$. Since $\cT(h)=\cT(h_I)\, \cT(h_c)$, we have
$P_{\cT(h)}=P_{\cT(h_I)}$.
In the reducible case, the quantities that are blindly identifiable are
$h_I$, $A_I$, up to one scalar indeterminacy
(assuming certain identifiability conditions for the burst length $M$
and the excitation modes in $A$ \cite{carvalho:Ident99}).

To get $h=T_c\, h_I$ from $h_I$, there are $N_c\m 1$ indeterminacies
(the coefficients of $h_c$). To get $A$ from $A_I=\cT(h_c)A$, there
are also $N_c\m 1$ indeterminacies
(modes of $A$ that are potentially coinciding with zeros of $\rmH_c(z)$;
alternatively, one needs $N_c\m 1$ "initial conditions" to get $A$ from
$A_I$, given $h_c$ (which was already needed to get $h$ from $h_I$)).
So there are $2N_c \m 1$ indeterminacies all in all and hence
$J_{\theta\theta}$
has $2N_c \m 1$ singularities.
Now,
\beq
J_{hh}(\theta) =  \dfrac{1}{\sv} \cA^H P^{\perp}_{\cT(h)}\cA
 = \dfrac{1}{\sv}  \cA^H P^{\perp}_{\cT(h_I)}\cA
\label{eq6.2.5.2}\eeq
has $N_c$ singularities. Indeed, an alternative decomposition
for $h$ is $h=T_I\, h_c$ where $T_I$ is block Toeplitz with
$[h_I^T\;\; 0_{1\times{}(N_c\m 1)m}]^T$ as first column.
Now consider $h^{'}=T_I\, h_c^{'}$ where $h_c^{'}$ is
arbitrary (not monic). Then $\cA\, h^{'} = \cT(h^{'})\, A =
\cT(h_I)\cT(h_c^{'})A$.
Hence $J_{hh}(\theta)h^{'}=0$ and $\mbox{null} \la J_{hh}(\theta)\ra =\range\la
T_I\ra$.
So we get:

\begin{result}
The CRB for estimating a reducible $h$ under the constraint
$T_I^{o H}\, h = T_I^{o H}\, h^o$ is
\beq
CRB_{C,h} = J_{hh}^+(\theta) = \sv(\cA^H P^{\perp}_{\cT(h)}\cA)^+ \; .
\label{eq6.2.5.3}\eeq
\end{result}

Note that this set of constraints implies in particular $h^{o H} h = h^{o
H}h^o$.
Note also that under these constraints, an estimate $\widehat{h}$ will not
necessarily be of the form $\widehat{h} = \widehat{T}_I\, \widehat{h}_c$
with $\widehat{h}_c$ equal to $h_c$ or not:
$\widehat{h}$ is not necessarily reducible.
Nevertheless, the constraints mentioned are sufficient to make
$h$ identifiable. Indeed, identifiability of $h$ with deterministic symbols
implies being able to determine $h$ from the noise--free signal.
If we do that with for instance the subspace fitting method, then the signal
subspace will be $\range\la \cT(h_I^o)\ra$. Subspace fitting will force
$\range\la\cT(\widehat{h})\ra\subset\range\la\cT(h_I^o)\ra$ which implies
$\widehat{\bmH}(z) = \bmH_I^o(z)\, \widehat{\rmH}_c(z)$. The application
of the constraints now leads to $\widehat{\rmH}_c(z) = \rmH_c^o(z)$
and hence $\widehat{h}=h^o$.

If we want the estimated channel to be reducible, then we have to apply
the explicit constraint $h = T_c^o\, h_I$, which can be reformulated as the
following implicit constraint: $\cK_{\theta_1} = P^{\perp}_{T_c^o}h = 0$.
It turns out that in this case of deterministic input symbols, we can
remain working in the complex domain, which we shall do.
So we get
$\dfrac{\partial \cK^H_{\theta_1}}{\partial \theta_1} = P^{\perp}_{T_c^o}$
and we can take $\cV_{\theta_1} = P_{T_c^o}$.
Hence, the constrained CRB for $h$ satisfying the constraints
$P^{\perp}_{T_c^o}h=0$ (compare to (32)) and $h^{o H} h = h^{o H}h^o$ is
\beq
CRB_{C,h} = (P_{T_c}\, \cA^H P^{\perp}_{\cT(h)}\cA\, P_{T_c})^+
= (P_{T_c}\, J_{hh}(\theta)\, P_{T_c})^+
\label{eq6.2.5.4}\eeq
which in general will be smaller than $J_{hh}^+(\theta)$ since more
a priori information is introduced (in the form of $m(N_c\m 1)\p 1$
constraints, compared to the minimal number of $N_c$ constraints to ensure
identifiability).

%%%%%%%%%%%%%%%%%%%%%%%%%%%%%%%%%%%%%%%%%%%%%%
\subsection{Gaussian Model}
%%%%%%%%%%%%%%%%%%%%%%%%%%%%%%%%%%%%%%%%%%%%%%
\label{secgaus}

In the Gaussian model, the estimation of $h$ is not decoupled from the
estimation of $\sv$ and the estimation parameter is 
$\theta = \lsb h^H \;\;\sv\rsb^H$.
Unlike in the deterministic model, as $J_{\theta\theta^*}\neq 0$,
we cannot treat the complex and real constellations together.
Identifiability here occurs from $C_{YY}(\theta)$ as $m_Y(\theta)\equiv 0$.
Blind identifiability in the Gaussian model does not require the channel
to be irreducible~\cite{carvalho:Ident99}.

\subsubsection{FIMs}

\paragraph{Circular Complex Symbol Constellation}

When the input constellation is complex, the FIM computation is based on the 
complex  probability  density function of $\bmY$:
\beq
\begin{array}{lcll}
\bmY\sim\cN(m_Y, C_{YY}), &\mbox{ with }& C_{YY}=\sa \cT(h)\cT^H(h)+\sv I, &
m_Y= 0 \, .
\end{array}
\label{C2eq71}
\eeq
Let $h_R=[\re(h^T)\; \im(h^T)]^T$ 
and $\overline{\theta}_R=[h^T_R \; \sv]^T$, the real  parameter vector.
As $J_{\theta\theta^*}$ is non zero, we cannot consider the complex CRB anymore:
the real FIM$\cJ_{\overline{\theta}_R \overline{\theta}_R}$
is determined via (\ref{C2eq7}) thanks to the quantities:
\beq
J_{\theta\theta}(i,j) =  
\trace\la C_{YY}^{-1}\lb  \dfrac{\partial C_{YY}}{\partial \theta_i^*}\rb 
C_{YY}^{-1}  \lb \dfrac{\partial C_{YY}}{\partial \theta_j^*} \rb^H  \ra
\label{C2eq72}
\eeq
\beq
J_{\theta \theta^*}(i,j)  = 
\trace\la C_{YY}^{-1}\lb  \dfrac{\partial C_{YY}}{\partial \theta_i^*}\rb 
C_{YY}^{-1}  \lb \dfrac{\partial C_{YY}}{\partial \theta_j^*} \rb  \ra
\label{C2eq73}
\eeq
\beq
\mbox{where: } \la
\begin{array}{l}
\dfrac{\partial C_{Y Y}}{\partial h^*_{i}}= 
\sa \cT(h)\cT^H \lb {\dfrac{\partial h}{\partial h^*_{i}}} \rb \\
\dfrac{\partial C_{YY}}{\partial \sigma_v^2}= \dfrac{1}{2}I \; .
\end{array}
\right. 
\label{C2eq74}
\eeq

\paragraph{Real Symbol Constellation}

When the input constellation is real, the FIM is:
\beq
\cJ_{\theta\theta}(i,j)= \dfrac{1}{2}\trace\la C_{YY}^{-1}\lb  \dfrac{\partial C_{YY}}{\partial \theta_i}\rb 
C_{YY}^{-1}  \lb \dfrac{\partial C_{YY}}{\partial \theta_j} \rb^T  \ra
\label{C2eq75}
\eeq
\beq
\la
\begin{array}{l}
\dfrac{\partial C_{Y Y}}{\partial h_{i}}= 
\sa \cT(h) \cT^T \lb {\dfrac{\partial h}{\partial h_{i}}} \rb 
+
\sa \cT \lb {\dfrac{\partial h}{\partial h_{i}}} \rb \cT^T(h)
\\
\dfrac{\partial C_{YY}}{\partial \sigma_v^2}= I \; .
\end{array}
\right.
\label{C2eq76}
\eeq

\subsubsection{FIM singularities}

\paragraph{Circular Complex Symbol Constellation}
Under the Gaussian blind identifiability conditions~\cite{carvalho:Ident99}, 
a complex channel $h$ is identifiable up to a phase factor. 
This corresponds to one singularity of the global FIM 
$\cJ_{\overline{\theta}_R \overline{\theta}_R}$:
\beq
\cJ_{\overline{\theta}_R\overline{\theta}_R} =
\lsb
\begin{array}{cc}
\cJ_{h_R h_R} & \cJ_{h_R \sv} \\
\cJ_{\sv h_R} & \cJ_{\sv \sv}
\end{array}
\rsb
\label{C2eq77}
\eeq 
 as well as of: 
\beq
\cJ_{h_R h_R}(\overline{\theta}_R) =  
\cJ_{h_R h_R} -
\cJ_{h_R \sv}
\lb \cJ_{\sv \sv}\rb^{-1} 
\cJ_{\sv h_R} \; .
\label{C2eq78}
\eeq
$\cJ_{h_R h_R}^{-1}(\overline{\theta}_R)$ would be the unconstrained CRB
for $h$ if its estimation were regular.
%It can be verified that $\cJ_{\overline{\theta}_R\overline{\theta}_R}$ 
%and $\cJ_{h_R h_R}(\overline{\theta}_R)$ have 
%the same number of null vectors, and can then be considered as equivalent from this
%point of view. 
%
The null space of $\cJ_{h_R h_R}(\overline{\theta}_R)$ is  spanned by 
\beq
{h}_{S}=\lsb -\im^T(h) \;\; \re^T(h) \rsb^T = h_{S_2} \; .
\label{C2eq79}
\eeq

\paragraph{Real Symbol Constellation}
The real FIM $\cJ_{\theta \theta}$ is regular under the identifiability
conditions, as well as $\cJ_{hh}(\theta)$.

\subsubsection{Equivalence between FIM regularity and  local identifiability}

\begin{thm}
The (complex or real) FIM is singular if and only if there exist a vector 
$h'$ (complex or real) and  a scalar $\sigma_v^{2 \,'}$ such that:
\beq
\sa \cT(h) \cT^H(h') + \sa\cT(h') \cT^H(h) +\sigma_v^{2 \,'} I = 0 \; .
\label{C2eq80}
\eeq
\label{thm16}
\end{thm}
\noindent
{\em Proof:}

\noindent
{\bf Complex case:}
The complex FIM matrix is singular if there exists a  $\overline{\theta}'_R
=\lsb \re^T(h') \;\; \im^T(h') \;\; \sigma_v^{2 \,'}\rsb^T$, such that:
\beq
\cJ_{\overline{\theta}_R \overline{\theta}_R} \overline{\theta}'_R=0 
\label{C2eq81}
\eeq
\beq
\;\; \Leftrightarrow \;\;
\lsb 
\begin{array}{ccc}
\lb\dfrac{\partial \Vec^T\{C_{YY}\}}{\partial h^*} \rb^T&
\lb\dfrac{\partial\Vec^T\{C_{YY}\}}{\partial h} \rb^T&
\lb\dfrac{\partial\Vec^T\{C_{YY}\}}{\partial \sv}\rb^T
\end{array}
\rsb
\lsb 
\begin{array}{c}
h'\\
h^{'\, *}\\
\sigma_v^{2 \,'}
\end{array}
\rsb
=0
\label{C2eq82}
\eeq
\beq
\;\;\Leftrightarrow \;\; 
\dsum_j  \lb \frac{\partial C_{YY}}{\partial h_j^*} \rb^H  h'_j
+
\dsum_j  \lb \frac{\partial C_{YY}}{\partial h_j^*} \rb  {h'_j}^* 
+
\dfrac{1}{2} \sigma_v^{2\,'} I = 0  \; .
\label{C2eq83}
\eeq
We have: 
$\ds{\frac{\partial C_{YY}}{\partial h_j^*} = \sa
\cT(h) \cT^H\lb{\frac{\partial h}{\partial h_i}} \rb}$
 and 
$\dsum_j \cT\lb{\frac{\partial h}{\partial h_i}}\rb h'_j=\cT(h')$,
then:
\beq
(\ref{C2eq81})
\;\; \Leftrightarrow \;\;
\sa \cT(h') \cT^H(h)  +  \sa \cT(h) \cT^H(h') + \dfrac{1}{2}\sigma_v^{2\,'} I =0
\; .
\label{C2eq85}
\eeq
which is equivalent to equation (\ref{C2eq80}) (with
$\frac{1}{2}\sigma_v^{2\,'} \rightarrow \sigma_v^{2\,'}$).

\hfill $\square$

\noindent
{\bf Real case:}
The real FIM matrix is singular if there exists a  ${\theta}' 
=\lsb {h'}^T \;\; \sigma_v^{2 \,'}\rsb^T$, such that:
\beq
\cJ_{{\theta}{\theta}} {\theta}'=0 
\label{C2eq86}
\eeq
\beq
%(\ref{C2eq86})
\;\; \Leftrightarrow \;\;
\lsb 
\begin{array}{cc}
\lb\dfrac{\partial\Vec^T\{C_{YY}\}}{\partial h} \rb^H&
\lb\dfrac{\partial\Vec^T\{C_{YY}\}}{\partial \sv}\rb^H
\end{array}
\rsb^H 
\lsb 
\begin{array}{c}
h'\\
\sigma_v^{2 \,'}
\end{array}
\rsb
=0
\label{C2eq87}
\eeq
\beq
\;\; \Leftrightarrow \;\;  
\dsum_j  \lb \frac{\partial C_{YY}}{\partial h_j} \rb^H  h'_j
+
\sigma_v^{2\,'}I  = 0 \; .
\label{C2eq88}
\eeq
We have 
$\ds{\frac{\partial C_{YY}}{\partial h_j} = \sa
\cT(h) \cT^H \lb{\frac{\partial h}{\partial h_i}} \rb + 
\sa \cT^H\lb{\frac{\partial h}{\partial h_i}}\rb \cT^H(h)}$.
Then:
\beq
(\ref{C2eq86})
\;\; \Leftrightarrow \;\;
\sa \cT(h') \cT^H(h)  +  \sa \cT(h) \cT^H(h') +  \sigma_v^{2\,'} I =0 \; .
\label{C2eq90}
\eeq

\hfill $\square$

\noindent
From (\ref{C2eq80}), we can deduce the following theorem.
\begin{thm}
The real/complex channel is locally blindly identifiable
%~\footnote{locally
%  identifiable up to a phase factor for a complex channel and strictly
%  identifiable for a real channel} 
if and only if the FIM is regular/1--singular. 
\end{thm}
Note that locally a complex channel is identifiable up to a continuous phase 
factor but a real channel is locally identifiable strictly speaking. 

\vspace{2mm}
\noindent 
{\em Proof:}
Assume that the FIM has a null vector $\theta'=[{h'}^T\;\; {\sigma_v^{2\,'}}]^T$ 
which in the complex channel case is non colinear to $h_S$.
Then theorem~\ref{thm16} says  that $\theta'$ satisfy (\ref{C2eq80}). 
Now, with $\eps>0$  arbitrarily small,
\beq
\begin{array}{lcl}
C_{YY}(\theta\p\eps \theta')\m C_{YY}(\theta) &\!\!\! = \!\!\!& 
\lb \sa \cT(h\p\eps h')\cT^H(h\p\eps h')\p\lb{\sigma_v^2}\p\eps\sigma_v^{2 \,'}\rb I\rb 
\m \lb \sa \cT(h)\cT^H(h) \p {\sigma_v^2} I  \rb\\
&\!\!\!=\!\!\! & \sa\cT(h)\cT^H(\eps h') \p \sa\cT(\eps h')\cT^H(h) \p \eps\; {\sigma_v^{2 \,'}} I
\p O(\eps^2) = O(\eps^2) \; .
  \end{array}
\label{C2eq92}
\eeq
This means that the covariance matrix is locally constant in 
the direction of $\theta'$ around $\theta$.
% along the direction of ${\bfH'}^\dagger(z)$.
%Then, along this direction $\bfH(z)$, a neighborhood of $\bfH(z)$ cannot
%be found where there is identifiability: there is no local identifiability.
%
Similarly to the  proof of theorem~\ref{deteq}, one can show that if the channel is 
identifiable, the FIM is regular or 1--singular.

\hfill $\square$

In \ref{C3app2}, we study the conditions on the characteristics of the 
channel to have local identifiability. The results are contained in the 
theorem below. The channel is assumed reducible: $\bfH(z) =
\bfH_I(z)\rmH(z)$. 
%$\Lu_I$ is the minimal length $M$ for
%which $\cT_M(h_I)$ has full--column rank;  ($\Lu_I \leq N_I\m 1$). 
\begin{thm}
The Gaussian FIM for a real/complex multichannel is regular/1--singular and the channel
 is locally blindly identifiable if:
\begin{enumerate}[(1)] 
\item $M \geq \max(\Lu_I +1,N_c-1)$,
\item the channel has no conjugate reciprocal zeros, \ie there exists
no $z_o \in \Rr/ \Cr$ such that $\bfH(z_o) = \bfH(1/z^*_o)=0$.
\end{enumerate}
\end{thm}

\noindent 
{\em Proof:} \ref{C3app2}.

\hfill $\square$

The no conjugate reciprocal zeros condition was also given in~\cite{zeng:sp197}, but for the 
real channel case only, without mentioning that the complex case is singular in any
case.
Remark that, in particular, the Gaussian FIM is regular if there are arbitrary 
zeros (not in conjugate reciprocal pairs) due to the fact that a minimum phase 
channel is identifiable (example of local identifiability).

The monochannel case is treated in~\ref{C3app1}: the results mentioned above 
for a multichannel are valid here also except that 
the noise variance $\sv$ cannot be identified, which 
results in an additional singularity of the FIM when the channel has no
conjugate reciprocal zeros (when the channel has 
conjugate reciprocal zeros, there is no additional singularity).

\subsubsection{Regularized Blind  CRBs}

\paragraph{Complex symbol constellation}

As in the deterministic case, we
need to define a regularized CRB, by introducing some a priori knowledge 
on the parameters, allowing us to determine  the ambiguous phase factor. 
We assume that the channel is (blindly) identifiable: we do not treat the
monochannel or conjugate reciprocal zeros.

The estimation of $h_R$ is considered under the constraint:
\beq
h_{S_2}^{o\,T} h_R =0
\label{C2eq93}
\eeq
which leads to  the constrained CRB for $h_R$:
\beq
CRB_{C,h_R} = \cJ_{h_{R}h_{R}}^+(\theta)\; .
\label{C2eq94}
\eeq 
This linear constraint does not allow to estimate the phase factor completely 
and a sign ambiguity is left but not reflected in the FIM singularities 
as it is a discrete  ambiguity.
For MSE computation purposes, the sign ambiguity can be resolved by requiring 
$h_R^{o\,T}h_R >0$, which together with (\ref{C2eq93}) can be stated as $h^{o\,H}h >0$.

\paragraph{Real symbol constellation}

No regularization is necessary and the CRB is $\cJ_{hh}^{-1}(\theta)$.
To compare the MSE for an estimator to this CRB, 
the knowledge of the right sign and right phase of the zeros
(\eg minimum phase in the reducible case)
should be used.

%All this study can be applied to the case of monochannels. The FIM will be
%regular except if one of the zeros is unitary.

%%%%%%%%%%%%%%%%%%%%%%%%%%%%%%%%%%%%%%%%%%%%%%
\section{Conclusions}
%%%%%%%%%%%%%%%%%%%%%%%%%%%%%%%%%%%%%%%%%%%%%%

We have provided a compact derivation of the CRB under parameter constraints.
In blind channel estimation under the deterministic or Gaussian symbol model,
the estimation problem has indeed to be augmented with constraints
to remove singularities. We have introduced the notion of minimal constraints
and shown how several intuitively attractive and hence popular constraint sets
lead simply to the pseudo inverse of the FIM as constrained CRB.
For the blind channel estimation problem, we have illustrated the
connection between local identifiability problems and FIM singularities.
In fact, if the symbol model would be a discrete alphabet constellation \cite{tsatsanis:spl98},
no more continuous ambiguities would persist and the unconstrained CRB would
typically exist. Deterministic and Gaussian symbol models have their
{\it raison d'\^etre} though, since they are typically less plagued with
local minima problems and hence can lead to estimates with which to initialize
methods based on discrete symbol alphabets.

%\newpage

%\appendix
\setcounter{section}{0}
\renewcommand{\thesection}{Appendix~\Alph{section}}
\renewcommand{\thesubsection}{\Alph{section}.\arabic{subsection}}
\renewcommand{\thesubsubsection}{\Alph{section}.\arabic{subsection}.\arabic{subsubsection}}

\section{}
\label{C3app1}
For a minimal number of independent constraints,
we prove that $CRB_C=\cJ_{\theta\theta}^+$ is the constrained CRB which 
gives the lowest value  for $\trace \la CRB_C \ra$ and is attained when
$\range\la\dfrac{\partial \cK_\theta}{\partial \theta}\ra=\mbox{null}(\cJ_{\theta\theta})$.

Let $\cK_\theta=0$ be an arbitrary set of  constraints on $\theta$; 
$\cV_\theta= \lb \frac{\partial \cK^T_\theta}{\partial\theta}\rb^\perp$ has full
column rank and we assume that $\cV^{T}_\theta \cJ_{\theta\theta} \cV_\theta$ is invertible.
The  corresponding constrained CRB  is:
\beq
CRB_C= \cV_\theta \lb \cV^{T}_\theta \cJ_{\theta\theta} \cV_\theta \rb^{-1} \cV^{T}_\theta
\label{C2eq95}
\eeq
Introduce the eigendecomposition of 
$\cJ_{\theta\theta} = S_1 {\Lambda_1} S_1^T + S_2\, 0\, S_2^T$. In general,
$\cV_\theta$ has components along $S_1$  and $S_2$:
$\cV_\theta= S_1 Q_1 + S_2 Q_2$. The fact that the constraints $\cK_\theta$
are independent and minimal in number implies that $Q_1$ is square and
invertible. Then we obtain
\beq
\begin{array}{rcl}
CRB_C & = & \cV_\theta \lb \cV^{T}_\theta S_1 \Lambda_1 {S_1}^T \cV_\theta \rb^{-1} \cV^{T}_\theta\\
 & = & \cV_\theta \lb Q_1^T \Lambda_1 Q_1 \rb^{-1} \cV^{T}_\theta\\
 & = & \cV_\theta  Q_1^{-1} \Lambda_1^{-1} Q_1^{-T}\cV^{T}_\theta\\
 & = & \lb {S_1} +  S_2 Q_2 Q_1^{-1} \rb \Lambda_1^{-1} \lb S_1 +  S_2 Q_2
 Q_1^{-1} \rb^T
\end{array}
\label{C2eq99b}
\eeq
The difference between the $CRB_C$ and $\cJ^+_{\theta\theta}={S_1}
\Lambda_1^{-1}  {S_1}^H$ may be indefinite in general, however:
\beq
\trace\lb{CRB_C}\rb = \trace\lb{\cJ^+_{\theta\theta}}\rb +
\trace\lb{Q_2 Q_1^{-1} \Lambda_1^{-1}Q _1^{-T} Q_2^T}\rb
\label{C2eq100}
\eeq
The second term is nonnegative, so 
$\trace\lb{CRB_C}\rb \geq \trace\lb{\cJ^+_{\theta\theta}}\rb$, with 
equality iff $Q_2=0$, 
\ie $\range\la\dfrac{\partial \cK_\theta^T}{\partial \theta}\ra
=\mbox{null}(\cJ_{\theta\theta})$ or
$\range\la \cV_{\theta}\ra = \range\la\cJ_{\theta\theta}\ra$.
In that case $CRB_C=\cJ_{\theta\theta}^+$.

\section{}
\label{C3app2}

%The singular vector of  
%$\cJ_{\overline{\theta}_R\overline{\theta}_R}=
%\lsb
%\begin{array}{cc}
%\cJ_{h_R h_R} & \cJ_{h_R \sv}\\
%\cJ_{\sv h_R} & \cJ_{\sv \sv}
%\end{array} 
%\rsb$
%corresponding to the phase ambiguity in $h$ is
%$\lsb -\im^T(h)\;\; \re^T(h) \;\; 0 \rsb^T = \lsb h^T_s \;\; 0 \rsb^T$ :
%\beq
%\cJ_{h_R h_R}h_s = 0 \mbox{ and } \cJ_{\sv h_R}h_s = 0  
%\label{C2eq118}
%\eeq
%In this appendix, we study the other singular vectors of 
%$\cJ_{\overline{\theta}_R\overline{\theta}_R}$, which are
%solutions $(h', \sigma_v^{2'})$ of the equation:
%\beq
%\sa \cT(h) \cT^H(h') + \sa\cT(h') \cT^H(h) +  \sigma_v^{2'} I=0
%\label{C2eq101}
%\eeq
%We first treat the monochannel case for a complex or real channel, 
%which allows us then to treat the multichannel case.

In this appendix, we study the solutions $(h', \sigma_v^{2'})$ of the equation:
\beq
\cT(h) \cT^H(h') + \cT(h') \cT^H(h) +  \sigma_v^{2'} I=0 \; .
\label{C2eq101}
\eeq
We first treat the monochannel case for a complex or real channel, 
which allows us to then treat the multichannel case.

\subsection{Complex Monochannel}

We assume 
that $M \geq N -1$; in this case, equation (\ref{C2eq101}) can be written 
in the $z$--domain as:
\beq
\sa\rmH(z) {\rmH'}^\dagger(z) + \sa\rmH'(z)\rmH^\dagger(z) + \sigma_v^{2\,'}= 0
\label{C2eq102}
\eeq
where $\rmH^\dagger(z) = \rmH^H(1/z^*)$.
Let's denote $p(z)=\rmH(z) {\rmH'}^\dagger(z)$, then:
\beq
(\ref{C2eq102})
\Leftrightarrow
p(z) + p^\dagger(z) + \sigma_v^{2\,'} =0 \; .
\label{C2eq103}
\eeq

\subsubsection{Solutions of the form $[* \;\; * \;\; 0]^T$: $\sigma_v^{2\,'}=0$.}

\beq
p(z) = \dsum_{i=-d_p}^{d'_p} \alpha_i z^{-i}
\mbox{ and }
p^\dagger(z) = \dsum_{i=-d_p'}^{d_p} \alpha^*_{-i} z^{-i}
\label{C2eq104}
\eeq
\beq
p(z) + p^\dagger(z) =0
\Rightarrow
d'_p=d_p 
\hspace{5mm} (\mbox{and }
\alpha_i = - \alpha^*_{-i}) \; .
\label{C2eq105}
\eeq
As $\rmH(z)$ and $\rmH^\dagger(z)$ are respectively causal and anticausal,
$\deg(\rmH(z)) = \deg(\rmH'(z))=N-1=d_p$. In the following, we assume that 
$\rmH(z)$ is monic.
Equation (\ref{C2eq103}) is also equivalent to:
\beq
p(z)=-p^\dagger(z) \; .
\label{C2eq106}
\eeq
From this equation, we can deduce that if $z_o$ is a zero of $p(z)$,
so is $1/z_o^*$, which implies that $p(z)$ is of the form:
\beq
p(z)= \alpha \dprod_{i=0}^{N_1-1} (1 - z_i z^{-1})(1 -z_i^* z) \;
\lsb (1- z^{-1}) (1+z)\rsb^{N_2} \; .
\label{C2eq107}
\eeq
where $N_1+N_2 = N-1$. 
We will differentiate the zeros that are equal to $1$ or $-1$:
$\{z_i\}_{i=1:N_1-1}$ are different from $1$ or $-1$.
$z$ equals $1$ or $-1$.

The number of singularities depends on the characteristics of the channel $\rmH(z)$
and namely the presence of conjugate reciprocal zeros. 

\begin{enumerate}[(1)]

   \item $\rmH(z)$ has no conjugate reciprocal zeros:

The $N-1$ zeros of $\rmH(z)$ are among the zeros of $p(z)$,
this implies that $p(z)$ has no zeros equal to $1$ or $-1$:
\beq
p(z) = \alpha \dprod_{i=0}^{N-2} (1 - z_i z^{-1})
 \dprod_{i=0}^{N-2} (1 -z_i^* z) = \rmH(z) {\rmH'}^\dagger(z)
\label{C2eq108}
\eeq
furthermore,  without loss of generality, we can assume that:
\beq
\rmH(z) = \dprod_{i=0}^{N-2}(1 - z_i z^{-1}) \; .
\label{C2eq109}
\eeq
In that case:
\beq
{\rmH'}^\dagger(z) = \alpha  \dprod_{i=0}^{N-1} (1 -z_i^* z) 
= \alpha {\rmH}^\dagger(z) 
\label{C2eq110}
\eeq
\beq
(\ref{C2eq103})
\Rightarrow
{\rmH'}(z) = j \rmH(z) \; .
\label{C2eq111}
\eeq
The FIM is 1-singular. Its null space is spanned by 
$\lsb -\im^T(h)\;\; \re^T(h)\rsb^T$.

    \item $\rmH(z)$ has 1 pair of conjugate reciprocal zeros: 
$(z_o, 1/z_o^*)$, $z_o \neq 1$, $z_o \neq -1$.  

Again, without loss of generality, we can assume that:
\beq
\rmH(z) = (1 - z_o z^{-1})(1 -z_o^* z) z^{-1} z_o^{-*}   
\underbrace{\dprod_{i=1}^{N-3} (1 - z_i z^{-1})}_{\rmH_1(z)} \; .
\label{C2eq112}
\eeq
There are 2 degrees of freedom in $\rmH'(z)$ coming from the fact that 
$\rmH'(z)$ can admit $1$ and $-1$ as zeros or not. Two possible choices 
for $\rmH'(z)$ are then:
\beq
\la
\begin{array}{l}
\rmH'(z) =  j (1 - z_{N-1} z^{-1}) (z^{-1} -z_{N-1}^*)z_{o} \; \rmH_1(z) \; ,\\
\rmH'(z) = (1-z^{-1})(z^{-1}-1)  z_o \; \rmH_1(z).
\end{array}
\right.
\label{C2eq113}
\eeq
The FIM has 2 singularities coming from the pair of conjugate reciprocal zeros,
and 1 singularity corresponding to $j \rmH(z)$.

\item $\rmH(z)$ has one zero equal to $1$ or $-1$.

We assume that this zero is equal $1$. $\rmH'(z)$ can be chosen as:
\beq
\rmH(z) = (1 - z^{-1})   
\underbrace{\dprod_{i=0}^{N-3} (1 - z_i z^{-1})}_{\rmH_1(z)}\, ,\;
\rmH'(z) = (1 + z^{-1})  {\rmH_1(z)}\; .
\label{C2eq115}
\eeq

\item $\rmH(z)$ has several conjugate reciprocal zeros: 

Then, to each pair of conjugate reciprocal zeros different from
$1$ and $-1$, correspond 2 singularities, and to each zero equal to 
$1$ or $-1$ corresponds 1 singularity.

\end{enumerate}

\subsubsection{Solutions of the form $[* \;\; * \;\; \sigma_v^{2\,'}]^T$: $\sigma_v^{2\,'}\neq 0$}

\begin{enumerate}[(1)]

 \item $\rmH(z)$ admits  conjugate reciprocal zeros $z_o$:
\beq
\underbrace{p(z_o)}_{=0} + \underbrace{p^\dagger(z_o)}_{=0} +\sigma_v^{2\,'} = 0
\Rightarrow
\sigma_v^{2\,'} = 0 \; .
\label{C2eq116}
\eeq
So there is no singular vector of the desired form in this case.

\item $\rmH(z)$ has no conjugate reciprocal zeros:

$\rmH(z)$ is of the form $\rmH(z) = \dprod_{i=0}^{N-2}(1 - z_i z^{-1})$. 
One can verify that 
\beq
\rmH'(z)= \dprod_{i=0}^{N-2}(1 + z_i z^{-1})
\eeq
is such that:
\beq
\sa\rmH(z)\rmH'(z)+\sa\rmH'(z)\rmH(z)=
\sa 2^{N\m 1}\dprod_{i=0}^{N-2}(1-\|z_i\|^2) \; .
\eeq
And so 
$H'(z)$ and $\sigma_v^{2\,'}=-\sa 2^{N\m 1}\dprod_{i=0}^{N-2}(1-\|z_i\|^2)$ 
verify (\ref{C2eq102}); and it can also be proved that this is the only 
singular vector due to the unidentifiability of $\sv$. 
It can also be verified that $\rmH'(z)$ is not a solution of (\ref{C2eq102}) if 
$\rmH(z)$ has conjugate reciprocal zeros. 

\end{enumerate}

\subsection{Real Monochannel}

\subsubsection{Solutions of the from $[* \;\; * \;\; 0]^T$: $\sigma_v^{2\,'}=0$.}

Similar reasonings apply here.

\begin{enumerate}[(1)]

  \item $\rmH(z)$ has no pair of conjugate reciprocal zeros:

$p(z) + p^{\dagger}(z)=0$ can only be satisfied by $p(z)\equiv 0$.
So the FIM is regular.

\item $\rmH(z)$ has 1 pair conjugate reciprocal zeros: 
$(z_o,1/z_o^*)$, $z_o \neq 1$, $z_o \neq -1$.  
\beq
\rmH(z) = (1 - z_o z^{-1})(1 -z_o z) z^{-1} z_o^{-1}   
\underbrace{\dprod_{i=1}^{N-3} (1 - z_i z^{-1})(1 - z_i z) }_{\rmH_1(z)} \; .
\label{C2eq119}
\eeq
There is now only 1 $\rmH'(z)$ possible (the first solution in (\ref{C2eq113})
is not valid here):
\beq
\rmH'(z) = (1-z^{-1})(1-z) z^{-1} z_o \rmH_1(z) \; .
\label{C2eq120}
\eeq
The FIM has 1 singularity.

\item $\rmH(z)$ has one zero equal to $1$ or $-1$.

We assume that this zero is $1$. $\rmH'(z)$ can be chosen as:
\beq
\rmH(z) = (1 - z^{-1})   
\underbrace{\dprod_{i=0}^{N-2} (1 - z_i z^{-1})}_{\rmH_1(z)}
\, ,\;
\rmH'(z) = (1 + z^{-1})  {\rmH_1(z)}\; .
\label{C2eq122}
\eeq

\item $\rmH(z)$ has several conjugate reciprocal zeros: 

Then, to each pair of conjugate reciprocal zeros different from $1$  and $-1$,
and to each zero equal to $1$ or $-1$ corresponds  1 singularity.

\end{enumerate}

\subsubsection{Solutions of the form $[* \;\; * \;\; \sigma_v^{2\,'}]^H$: $\sigma_v^{2\,'}\neq 0$}

The same singularity as in the complex case, due to the inidentifiability
of $\sigma_v^2$, appears (except again if the channel $\rmH(z)$ has conjugate
reciprocal zeros).

\subsection{Multichannel}

Assume now that $\bfH(z)$ is a true multichannel, possibly reducible:
\beq
\bfH(z) = \bfH_I(z) \rmH_c(z) \; . 
\label{C2eq124}
\eeq
As for the monochannel case, we search first the solutions of the 
form: $\lsb \re^T(h') \;\; \im^T(h') \;\; 0 \rsb^T$. Then $h'$ verifies:
\beq
 \cT(h)\cT^H(h') +  \cT(h') \cT^H(h) = 0 \; .
\label{C2eq125}
\eeq
The burst length is assumed to be $M \geq \Lu+1$ which can be lower than
$N-1$ (so the transposition to the z-domain is not as convenient as in 
the monochannel case).
The previous equation implies that $\cT(h')$ should have for effect to 
reduce the previous quantity to at least the same rank as $\cT(h)$. So:
\beq
\range\{\cT(h')\} \subset \range\{\cT(h)\}
\label{C2eq126}
\eeq
which implies (see e.g.\   Appendix B in \cite{carvalho:Ident99}):
%%app
\beq
\bfH'(z) =\bfH_I(z)\rmH'_c(z)
\label{C2eq127}
\eeq
\beq
\la
\begin{array}{rcl}
\bfH'(z) & = & \bfH_I(z) \rmH'_c(z) \\
\bfH(z) & = & \bfH_I(z) \rmH_c(z) 
\end{array}
\right.
\hspace{1mm}
\Rightarrow
\hspace{1mm}
\cT(h_I) \cT(h_c) \cT^H(h_c')\cT^H(h_I) + \cT(h_I) \cT(h_c')\cT^H(h_c)
\cT^H(h_I)=0 \; .
\label{C2eq128}
\eeq
As $\cT(h_I)$ is full column-rank, 
\beq
(\ref{C2eq128})
\Leftrightarrow
 \cT(h_c) \cT^H(h_c')+\cT(h_c')\cT^H(h_c)  = 0 \; .
\label{C2eq129}
\eeq
As $\cT(h_c)$ is of length at least $N_c\m 1$ according to the identifiability
conditions,  (\ref{C2eq129}) implies:
\beq
\rmH_c(z) {\rmH'}^\dagger_c(z) + {\rmH'}^\dagger_c(z)\rmH_c(z) = 0
\label{C2eq130}
\eeq
which leads to the monochannel case treated previously.

As for the solutions of the form $\lsb *\;\; \sigma_v^{2'} \rsb^T$,
$\sigma_v^{2'}\neq 0$, there are none in this case ($\sigma_v^2$ is
identifiable in any case).

%\newpage

%\bibliography{../refsIEEE}

\begin{thebibliography}{10}\itemsep 0mm

\bibitem{carvalho:Ident99}
E.~de~Carvalho and D.T.M.\ Slock,
\newblock ``{Blind and Semi--Blind FIR Multichannel Estimation: Identifiability
  Conditions},''
\newblock Submitted to IEEE Transactions on Signal Processing.

\bibitem{Moulines:sp95}
E.\ Moulines, P.\ Duhamel, J.F.\ Cardoso, and S.\ Mayrargue,
\newblock ``{Subspace Methods for the Blind Identification of Multichannel FIR
  filters},''
\newblock {\em IEEE Transactions on Signal Processing}, vol. 43, no. 2, pp.
  516--526, Feb. 1995.

\bibitem{carvalho:SBCRB99}
E.~de~Carvalho and D.T.M.\ Slock,
\newblock ``{Blind and Semi--Blind FIR Multichannel Estimation: Cram\'er--Rao
  Bounds},''
\newblock Submitted to IEEE Transactions on Signal Processing.

\bibitem{Hero:it90}
J.D.\ Gorman and A.O.\ Hero,
\newblock ``{Lower Bounds for Parametric Estimation with Constraints},''
\newblock {\em IEEE Transactions on Information Theory}, vol. 26, no. 6, pp.
  1285--1301, Nov. 1990.

\bibitem{AbedMeraim:sp97}
K.\ Abed~Meraim, E.\ Moulines, and P.\ Loubaton,
\newblock ``{Prediction Error Method for Second-Order Blind Identification},''
\newblock {\em IEEE Transactions on Signal Processing}, vol. 45, no. 3, pp.
  694--705, March 1997.

\bibitem{giannakis:sp95}
G.B.\ Giannakis and S.D.\ Halford,
\newblock ``{Asymptotically Optimal Blind Fractionally-Spaced Channel
  Estimation and Performance Analysis},''
\newblock {\em {IEEE Transactions on Signal Processing}}, vol. 45, no. 7, pp.
  1815--1830, July 1997.

\bibitem{Kay:book}
S.M.\ Kay,
\newblock {\em {Fundamentals of Statistical Signal Processing Estimation
  Theory}},
\newblock Prentice-Hall, Englewood Cliffs, NJ, 1993.

\bibitem{Caines:book}
Peter~E.\ Caines,
\newblock {\em Linear Stochastic Systems},
\newblock John Wiley \& Sons, 1988.

\bibitem{hochwald:cssp97}
B.\ Hochwald and A.\ Nehorai,
\newblock ``{On Identifiability and Information-Regularity in Parameterized
  Normal Distributions},''
\newblock {\em Circuits, Systems ans Signal Processing}, vol. 16, no. 1, 1997.

\bibitem{marzetta:sp93}
T.\ Marzetta,
\newblock ``{On Simple Derivation of the Constrained Multiple Parameter
  Cram\'er--Rao Bound},''
\newblock {\em IEEE Transactions on Signal Processing}, vol. 41, no. 6, pp.
  2247--2249, June 1993.

\bibitem{stoica:spl98}
P.\ Stoica and B.C.\ Ng,
\newblock ``{On the Cram\'er--Rao Bound Under Parametric Constraints},''
\newblock {\em IEEE Signal Processing Letters}, vol. 5, no. 7, pp. 177--179,
  July 1998.

\bibitem{carvalho:spawc97}
E.\ de~Carvalho and D.T.M.\ Slock,
\newblock ``{Cram\'er-Rao Bounds for Semi-blind, Blind and Training Sequence
  based Channel Estimation},''
\newblock in {\em Proc.\ SPAWC 97 Conf.}, Paris, France, April 1997.

\bibitem{Hua:sp96}
Y.\ Hua,
\newblock ``{Fast Maximum Likelihood for Blind Identification of Multiple FIR
  Channels},''
\newblock {\em IEEE Transactions on Signal Processing}, vol. 44, no. 3, pp.
  661--672, March 1996.

\bibitem{HuaWax:sp96}
Y.\ Hua and M.\ Wax,
\newblock ``{Strict Identifiability of Multiple FIR Channels Driven by an
  Unknown Arbitrary Sequence},''
\newblock {\em IEEE Transactions on Signal Processing}, vol. 44, no. 3, pp.
  756--759, March 1996.

\bibitem{Bapat:thesis}
J.L.\ Bapat,
\newblock {\em Partially Blind Identification of FIR Channels for QAM Signals},
\newblock Ph.D. thesis, Pennsylvania State University, Aug. 1996.

\bibitem{morgan:spl98}
D.R.\ Morgan, J.\ Benesty, and M.M.\ Sondhi,
\newblock ``{On the Evaluation of Estimated Impulse Responses},''
\newblock {\em IEEE Signal Processing Letters}, vol. 5, no. 7, pp. 174--176,
  July 1998.

\bibitem{zeng:sp197}
H.H.\ Zeng and L.\ Tong,
\newblock ``{Blind Channel Estimation Using the Second--Order Statistics:
  Asymptotic Performance and Limitations},''
\newblock {\em IEEE Transactions on Signal Processing}, vol. 45, no. 8, pp.
  2060--2071, Aug. 1997.

\bibitem{tsatsanis:spl98}
H.A.\ Cirpan and M.K.\ Tsatsanis,
\newblock ``{Stochatic Maximum--Likelihood Methods for Semi--Blind Channel
  Estimation},''
\newblock {\em IEEE Signal Processing Letters}, vol. 5, no. 1, pp. 21--24, Jan.
  1998.

\end{thebibliography}

\end{document}